\documentclass[english,reprint,amsmath,amssymb,aps,superscriptaddress,eqsecnum,floatfix]{revtex4-1}
\usepackage[T1]{fontenc}
\usepackage[latin9]{inputenc}
\usepackage[dvipsnames]{xcolor}
\usepackage{babel}
\usepackage{units}
\usepackage{bm}
\usepackage{amsmath}
\usepackage{graphicx}
\usepackage[export]{adjustbox} 
\usepackage[unicode=true,pdfusetitle,
 bookmarks=true,bookmarksnumbered=false,bookmarksopen=false,
 breaklinks=false,pdfborder={0 0 1},backref=false,colorlinks=true]
 {hyperref}
\usepackage[caption=false]{subfig}

\makeatletter
\usepackage{babel}

\usepackage{babel}
\usepackage{multirow}
\usepackage{bm}

\newcommand{\bk}{{\bf k}}
\newcommand{\bP}{{\bf P}}
\newcommand{\bp}{{\bf p}}

\newcommand{\bR}{{\bf R}}
\newcommand{\br}{{\bf r}}

\newcommand{\bq}{{\bf q}}
\newcommand{\bv}{{\bf v}}
\newcommand{\bzeta}{{\boldsymbol\zeta}}

\newcommand{\im}{\text{Im}}
\newcommand{\ph}{\text{ph}}
\newcommand{\dis}{\text{dis}}

\newcommand{\imp}{\text{imp}}
\newcommand{\ex}{\text{ex}}
\newcommand{\ch}{\text{ch}}

\newcommand{\hA}{{\hat A}}
\newcommand{\hB}{{\hat B}}
\newcommand{\hP}{{\hat P}}

\newcommand{\hPsi}{{\hat \Psi}}

\newcommand{\prep}{\text{prep}}

\newcommand{\Average}[1]{\left< #1 \right>}

\makeatother

\begin{document}

\title{Chaos and the dynamics of information in dissipative electronic systems}

\author{Markus J. Klug}
\email{markus.klug@kit.edu}

\selectlanguage{english}%

\affiliation{Institut für Theorie der Kondensierten Materie, Karlsruher Institut
für Technologie, 76131 Karlsruhe, Germany}
\affiliation{Physics Department, University of California, Santa Cruz, CA 95064, USA}

\author{Sergey V. Syzranov}

\affiliation{Physics Department, University of California, Santa Cruz, CA 95064, USA}
\email{syzranov@ucsc.edu}

\date{\today}
\begin{abstract}
The dynamics of chaotic systems are, by definition, exponentially sensitive to the initial
conditions and may appear rather random.
In this work, we explore relations between the chaotic dynamics of an observable   
and the dynamics of information (entropy) contained in this observable, focussing on a disordered metal 
coupled to a dissipative, e.g. phononic, bath. 
The chaotic dynamics is characterised
by Lyapunov exponents $\lambda$, the rates of growth of 
out-of-time order correlators (OTOCs), quantities of the form $\langle[\hat A(t),\hat B(0)]^{2}\rangle\propto\exp(2\lambda t)$, where $\hat A$
and $\hB$ are the operators of, e.g.,
the total current of electrons in a metallic quantum dot.
We demonstrate that the Lyapunov exponent $\lambda$ matches
the rate of decay of information stored in the observable $\langle \hat A(t)\rangle$
after applying a small perturbation with a small classical uncertainty.
This relation suggests a way to measure Lyapunov exponents in experiment.
We compute both the Lyapunov exponent and the rate of decay of information
microscopically in a disordered metal in the presence of a bosonic bath, which may, in particular, represent interactions in the system.
For a sufficiently short range of the correlations in the bath, the exponent has the form
$\lambda=\lambda_{0}-1/\tau$, where $\lambda_{0}$ is the (temperature-independent)
Lyapunov exponent in the absence of the bath
and $1/\tau$ is the inelastic scattering rate. Our results demonstrate also the existence of a transition
between chaotic and non-chaotic behaviour at $\lambda_{0}=1/\tau$, which may be triggered, e.g.,
by changing the temperature of the bath. 
\end{abstract}
\maketitle

\section{Introduction}

Classical chaotic systems are systems whose dynamics are exponentially sensitive to the initial conditions.
Chaotic dynamics is rather ubiquitous; it occurs in huge yet diverse classes of systems,
ranging from black holes~\cite{Maldacena:bound} to disordered metals~\cite{LarkinOvchnnikov}
and nanoscale quantum dots~\cite{Nakamura:book}. Due to the extreme sensitivity to the initial conditions,
chaotic dynamics may appear very random and their accurate description is, thus, extremely challenging.

While chaotic  behaviour in classical systems may be defined and quantified straightforwardly through 
exponential divergence of close trajectories of the system, describing and even defining chaotic 
dynamics in the case of a {\it quantum} system is more delicate.
A quantum system may be called chaotic if it has a classical chaotic limit.
However, a broad class of quantum systems, such as spin systems 
or Hubbard-type models,
do not even have the classical limit, while exhibiting evolution rather sensitive to external perturbations. Over several decades, the Wigner-Dyson statistics (see, e.g., Ref.~\cite{Efetov:book} and referenced therein)
of a system's 
energy levels has been used as a definition of quantum chaos. While this definition
 is consistent with the 
existence of classical chaotic dynamics in systems like disordered metals~\cite{LarkinOvchnnikov},
 the connection between
the level statistics and dynamics for an arbitrary system still remains to be investigated thoroughly.

Another characteristic of quantum chaotic behaviour, which has been in the focus of many studies 
of chaos since several years ago, is an out-of-time-order correlator (OTOC) 
of the form
\begin{align}
F\left(t\right)=-\langle\big[\hA(t),\hB(0)\big]^{2}\rangle, 
\label{eq:otoc}
\end{align}
where $\hA$ and $\hB$ are operators acting on the system's states,
 and the averaging is carried out with respect to the equilibrium
state of the system. 
This characteristic was first proposed~\cite{LarkinOvchnnikov} almost half a century ago 
in the context of non-interacting particles scattered off randomly located impurities, but
has come into focus of researchers' attention several years ago, causing an explosion of research 
activity on chaotic behaviour of  
both interacting and non-interacting quantum systems (see, e.g., Refs.~
\cite{Maldacena:bound,Stanford:firstOTOC,RozenbaumGalitski:rotor,AleinerFaoroIoffe,KlugSchmalian,BagretsAltlandaKamenev,PatelSachdev,
	PatelChowdhurrySachdevSwingle,WermanKivelsonBerg,RozenbaumGalitski:statistics,LiaoGalitski,RozenbaumBunimovichGalitski}).
The correlator \eqref{eq:otoc} describes the sensitivity of the operator $\hA(t)$ to a perturbation proportional 
to $\hB$. For a disordered metal, for example, the OTOC of momentum operators grows exponentially $\propto e^{2\lambda t}$,
where the exponent $\lambda$ matches~\cite{Syzranov:ChaosTransition} that of the divergence of classical trajectories~\cite{LarkinOvchnnikov},
but, unlike the case of a classical case, the growth persists only in a finite interval of time,
$\tau_{0} \ll t \ll t_E$, where $\tau_0$ is the elastic scattering time and $t_E$
is known as the Ehrenfest time, the characteristic time at which the crossover between classical and quantum dynamics occurs~\cite{BermanZaslavsky:Ehrenfest,LarkinOvchnnikov}.

Strictly speaking, OTOCs~\eqref{eq:otoc} are not experimentally measurable quantities, unless the experiment
involves effective time-reversal operations, such as spin-echo techniques~\cite{Slichter:book}
or reversing the sign of the Hamiltonian~\cite{GarttnerRey:ionOTOC,Knap:2copies} or using the second
copy of the system~\cite{Knap:2copies,Yao:2copies} or a control qubit~\cite{Swingle:measurement,ZhuHafeziGrover:measureClock,Danshita:SYKmeasurement}.   
Measurements of OTOCs have been limited so far to specific types of non-chaotic systems 
\cite{Li:NMRmeas,GarttnerRey:ionOTOC} of spins and trapped ultracold particles, yet
exponential growth of OTOCs still remains to be observed experimentally.

The concept of quantum chaos is related closely 
to the concept of quantum information scrambling, i.e. the process
of spreading of information in the system. Indeed, in the case of local operators,
correlators of the form~\eqref{eq:otoc} characterise the spreading of operators in space.
The exponential growth of an OTOC~\eqref{eq:otoc} is believed to cease when the information, stored initially
in the perturbation $\hA(0)$, is dispersed homogeneously
around the system and can no longer be extracted by local measurements.
The phenomenology of chaotic dynamics applies ubiquitously to an enormous variety of 
chaotic systems, ranging from black holes to mesoscopic structures and arrays of spins, which 
allows one to study it, in particular, by means of abstract toy models displaying chaotic dynamics,
such as Sachdev-Ye-Kitaev model~\cite{PatelChowdhurrySachdevSwingle}.

In this paper, our purpose is to investigate the chaotic dynamics in a quantum system
and the dynamics of information in it, as well as to establish connections between them.
We focus on the model of a disordered metal coupled to an external bath, possibly in the presence
of electron-electron interactions.
We study analytically the dynamics of both OTOCs
and the amount of information stored in the system's observables and analyse relations
between them.

\begin{figure}
	\centering
	\includegraphics[width=0.7\linewidth]{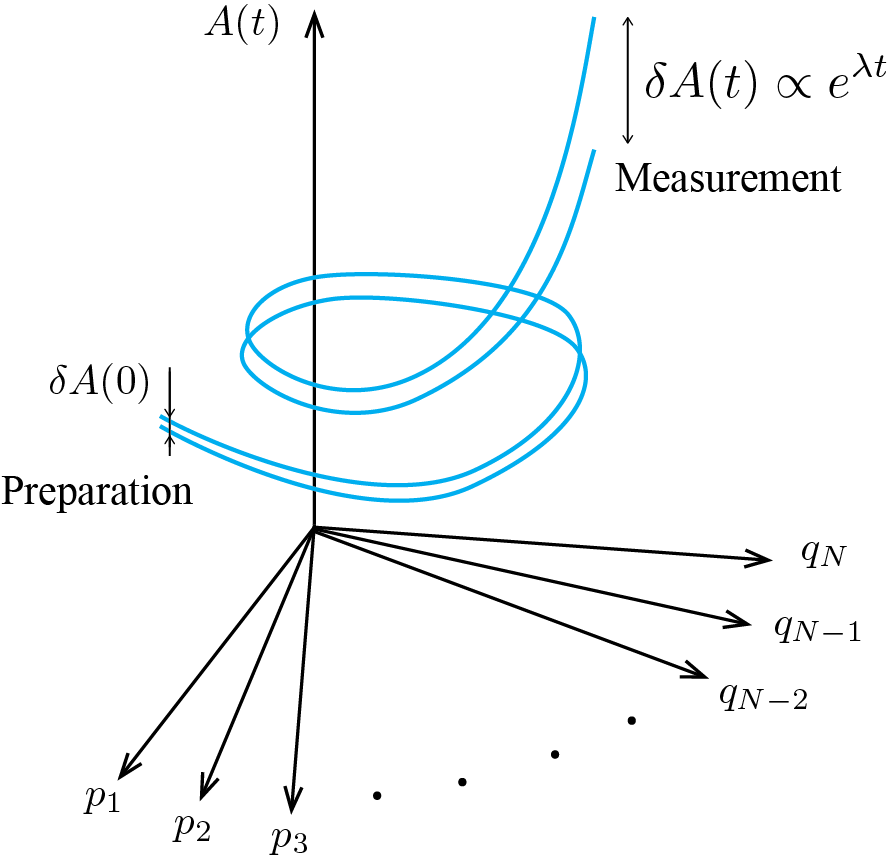}
	\caption{\label{fig:phasetrajectory}
	(Colour online) Phase trajectories of a closed classical chaotic system in the extended space of the observable
	$A(t)$ and phase space. $p_1, \ldots, p_N$ and $q_1,\ldots q_N$ are the 
	generalised momenta and coordinates.
	}	
\end{figure}

These relations are almost self-evident in the case of a closed system which has a classical limit.
Indeed, if the OTOC~\eqref{eq:otoc} in that system grows exponentially $\propto \exp(2\lambda t)$, 
this growth comes from the exponential divergence between close classical trajectories in the phase space
of the system with the Lyapunov exponent $\lambda$, as shown in Fig.~\ref{fig:phasetrajectory}.
Strictly speaking, the system may have multiple Lyapunov exponents, but for sufficiently 
generic operators $\hA$ and $\hB$ the correlator may be expected to grow with the maximum exponent
or the exponent having the biggest density in the spectrum of exponents. 
Let us assume that the initial state of the system has a small classical uncertainty, for example,
due to an imperfect preparation procedure, which leads also to a small uncertainty 
$\delta A(0)$ of the observable $A(0)=\langle \hA(0)\rangle $. Then the (linear) size of the 
uncertainty of the state of the system in phase space will grow exponentially in time with the Lyapunov exponent $\lambda$,
and so will the uncertainty $\delta A(t)$
of the observable $A(t)=\langle\hA(t)\rangle $, so long as this observable is sufficiently generic.

We define the amount of information [measured in units $(1/\ln 2)$th of a bit] contained in the quantity $A(t)$ as
\cite{Preskill:book} 
\footnote{For simplicity, instead of the logarithm to base 2, we choose the natural logarithm in the definition of information. 
The amount of information is therefore measured in units of bits times a multiplicative constant of ($ 1/\ln 2 $)}
\begin{equation}
I_{A}\left(t\right)=\text{const}+\int dA\,\rho (A,t) \ln\rho (A,t),
\label{eq:def-information}
\end{equation}
where $\rho(A,t)$ is the probability density of the quantity $A(t)$ and $\text{const}$ is an arbitrary constant. 
The amount of information \eqref{eq:def-information} is infinite when the quantity $A$ is known
exactly and reaches minimum when $A$ is distributed uniformly in the interval of values it may take.
The quantity $-I_A$, with $I_A$ given by Eq.~\eqref{eq:def-information}, is known also, up to a constant, as
the Shannon entropy \cite{Preskill:book} of the quantity $A(t)$. 
The order of magnitude of the probability density of $A(t)$ may be estimated as 
$\rho(A,t)\sim1/\delta A(t)\propto e^{-\lambda t}$ in the interval of the width of order $\delta A(t)$
near the mean value of $A(t)$. Using Eq.~\eqref{eq:def-information},
this suggests
that the amount of information associated with variable $A(t)$ depends on time as
$I_A(t)=\text{const}-\lambda^\prime t$, where the rate $\lambda^\prime$
is on the order of the Lyapunov exponent $\lambda$.
Thus, the rate of change information~(\ref{eq:def-information})
may be expected to
match, at least within the order of magnitude, the (maximum) Lyapunov exponent.

The relations between
chaotic dynamics, out-of-time-order correlators and the dynamics of information may be more complicated and 
require further investigation 
for open systems, i.e. coupled to a dissipative environment,
and for quantum systems, which may not have the classical limit.
This motivates us to study in this paper out-of-time-order correlators and the dynamics of information
in a system of electrons in a disordered metal coupled to a bath of neutral excitations, such as phonons 
or plasmons. In the absence of the coupling and for sufficiently large impurities, that system is known 
to be chaotic~\cite{LarkinOvchnnikov} (see also Ref.~\cite{Syzranov:ChaosTransition}) and have a well defined
(chaotic) classical limit. 
Coupling to a dissipative bath makes the dynamics
quantum and, for sufficiently strong coupling, may convert chaotic dynamics to non-chaotic.

This paper is organised as follows. In Sec.~\ref{sec:results} we present a summary
of our results. Sec.~\ref{sec:SingleParticle} deals with chaos and the dynamics of information
of a single particle in 
a disordered metal or a chaotic billiard.  
In Sec.~\ref{sec:Model} we introduce the microscopic model
of a disordered model in the presence of a phonon bath and/or interactions.
In Sec. \ref{sec:Out-of-time-order-correlator}, we evaluate the OTOC
of the form (\ref{eq:otoc}) to determine the Lyapunov exponent
$\lambda$ in this model. 
 Subsequently, the
concept of information associated with the system's total momentum
and its dynamics are discussed in Sec. \ref{sec:Information}. We
conclude in Sec.~\ref{sec:Conclusion}.


\section{Summary of results}
\label{sec:results}

We characterise the chaotic dynamics of a disordered metal coupled to a bath of neutral excitations by its
Lyapunov exponents, i.e. the rates $\lambda$ of growth 
 of out-of-time-order correlators of the form~\eqref{eq:otoc}.
In principle, the growth rate of an OTOC may depend on the choice of the operators $\hA$ and $\hB$,
and a generic system with quenched disorder should be expected to have a continuous spectrum of Lyapunov exponents.
However, an arbitrary pair of operators may be projected onto the operators which lead to the growth with the maximum exponent
or an exponent having a very large density in the spectrum of exponents,
which is why broad classes of similar operators $\hA$ and $\hB$ may be expected to lead to the exponential
growth with the same 
rate. 
A number of studies (see, for example, Refs.~\cite{Stanford:firstOTOC,KlugSchmalian,PatelChowdhurrySachdevSwingle,WermanKivelsonBerg,LiaoGalitski}) of out-of-time-order correlators 
 dealt with the creation $\hPsi^\dagger$ and annihilation $\hPsi$ operators 
of particles, for the sake of simplicity of calculations. However, correlators of observables, such as densities and currents, are even in creation and annihilation operators and, strictly speaking, 
may correspond to different Lyapunov exponents. In this paper, we consider the correlators
of the operators $\hA=\hB=\hP_z$ of the total momentum of electrons in a disordered metal coupled to 
a bath. We expect, however, that our results for the Lyapunov exponents  
apply for a broad class of operators, such as currents or densities of electrons (or their combinations).

The dissipative bath, to which the electrons under considerations are coupled, is represented 
by neutral excitations, hereinafter referred to as phonons. However, the role of
of the bath may be played by neutral excitations of other nature, e.g. magnons or plasmons.
Because these excitations may mediate interactions between electrons, we expect our results
to hold qualitatively for a systems of interacting electrons (in the many-body-delocalised
phase~\cite{BAA,NandkishoreHuse:review,AltmanVosk:review}) in the absence of a bath
as well as to non-interacting electrons coupled to a bath. 

{\it Lyapunov exponent.}
By evaluating the OTOC~(\ref{eq:otoc}) of the operators $\hP_z$ of the total momentum, we find that 
for a sufficiently short range of correlations in the phonon bath,
the Lyapunov exponent is given by
\begin{equation}
\label{eq:lya-exp}
\lambda=\lambda_{0}-{1}/{\tau},
\end{equation}
where $\lambda_0$ is the ``single-particle'' Lyapunov exponent, i.e. the rate
of exponential divergence between close classical trajectories~\cite{LarkinOvchnnikov} in the absence of the bath, and $\tau$ is the time of inelastic scattering of electrons by phonons.
The existence of the exponential growth of the OTOC requires that the impurities in the metal 
be sufficiently big, with the characteristic size $a$ exceeding $\left(\lambda_{F} v_F \tau_0\right)^\frac{1}{2}$,
where $\lambda_{F}$ and $v_F$ are the Fermi wavelength and velocity and $\tau_0$ is the elastic
scattering time. 
The same condition ensures the existence of chaotic behaviour for the single-particle 
problem~\cite{AleinerLarkin:longWL}. The chaotic behaviour, described by the 
exponent \eqref{eq:lya-exp}, persists in a finite interval of time $t_{\ph}\ll t\ll t_E$,
where 
\begin{equation}
t_{\ph}=\lambda_{0}^{-1}\ln (\xi_\text{ph}/ \lambda_{F}),
\label{eq:tph}
\end{equation}
with $\xi_\text{ph}$ being the correlations length of the phonon bath, and
\begin{equation}
t_{E}=\lambda_{0}^{-1}\ln (a/ \lambda_{F}) 
\end{equation}
is the Ehrenfest time~\cite{BermanZaslavsky:Ehrenfest,LarkinOvchnnikov}, which characterises
the crossover between classical and quantum dynamics of electrons 
in the non-interacting problem, i.e. in the absence of coupling to the bath.
At shorter times, $\tau_0 \ll t\ll t_\ph$, the OTOC also exhibits exponential growth, but with a different Lyapunov
exponent. In this paper, however, we focus on longer times, $t_\ph \ll t\ll t_E$, associated
with the Lyapunov exponent~\eqref{eq:lya-exp}.

Equation~\eqref{eq:lya-exp} suggest the existence of a transition between the regime of chaotic dynamics ($\lambda>0$), with the exponential
growth of the OTOC, and the regime of non-chaotic dynamics ($\lambda<0$), where the OTOC decays exponentially.
The transition may be triggered by changing the amount of disorder in the system or, e.g., the temperature, which affects 
the rate $1/\tau$ of inelastic scattering. This transition between chaotic and non-chaotic dynamics
is analogous to a similar transition predicted in Ref.~\cite{Syzranov:ChaosTransition} for a system of interacting
electrons in a disordered metal.

{\it Information dynamics.} We study also the relation of the chaotic dynamics, characterised by the
exponent~\eqref{eq:lya-exp}, to the dynamics of information associated with the observable 
$\hA$ which enters the out-of-time-order correlator. As a specific example,
we focus on the operator $\hA=\hP_z$ of the total momentum of electrons, but expect that our results hold as well
for a broad class of generic operators such as electron densities, total spins, currents or their combinations.

We assume that at time $t=0$ the distribution of the system is 
perturbed slightly, and that the perturbation has a small (classical) uncertainty and a short correlation
range in phase space.
This uncertainty 
makes the quantum-mechanical average $\langle \hA(t) \rangle $ of the observable $\hA$ a random function.
The amount of information associated with the quantity $A(t)=\langle \hA(t) \rangle $ changes with time.   
We demonstrate that in the case of short-range interactions the information changes with the rate
\begin{equation}
\tfrac{d}{dt}I_{A}\left(t\right)= -\lambda
\end{equation}
given by the Lyapunov exponent Eq.~(\ref{eq:lya-exp}). 
Thus, the externally induced fluctuations of an observable 
contain information about the system's chaotic dynamics.

When $\lambda_{0}> 1/\tau$, the system is chaotic, according to the definition of chaos by means of OTOCs, and
the uncertainty of the variable $A(t)$ grows. When $\lambda_{0}<1/\tau$, the strong coupling between the electrons and 
the bath leads to a fast relaxation of the system's state to equilibrium, which makes the observable $A(t)$ more 
certain, and the amount of information thus increases.
The established relation between OTOCs and the dynamics of information may be used, in principle, as a way of measuring 
Lyapunov exponents by monitoring the fluctuations of observables as a function of time. 
Several possible measurement schemes will be discussed in Sec.~\ref{sec:Conclusion}.


\section{Information dynamics for a single particle in a billiard}

\label{sec:SingleParticle}

Before describing chaotic dynamics in a many-body system in the
presence of a random potential and a dissipative bath, we 
use a single-particle example to illustrate the connection between 
chaotic behaviour and the dynamics of information contained in macroscopic observables.
 In what immediately follows, we consider the dynamics of a classical
particle scattered elastically
off randomly located impurities or, equivalently, the walls of a chaotic billiard. 

We assume that the initial momentum of the particle is not known exactly, but  
has a small uncertainty $\delta \bP_0$ relative to a certain vector $\bP_0$.
This (classical) uncertainty of the momentum may come, for example, from an imperfect
procedure of preparing the state of the particle
leading to (classical) randomness in the initial parameters of the system. 
The vector $\delta \bP _0$ of the uncertainty of the particle's momentum is perpendicular 
to the vector $\bP_0$ and has zero average, $\langle \delta \bP _0\rangle_\prep=0$,
where $\langle \ldots\rangle_\prep$ is our convention for averaging over
the procedure of preparing the initial state.
For simplicity, we assume also that the initial position
$\bR_0$ of the particle is known exactly, however, the analysis below may be generalised to the case where the 
initial position also has some uncertainty.

\begin{figure}[t]
	\centering{}\includegraphics[width=0.95\columnwidth]{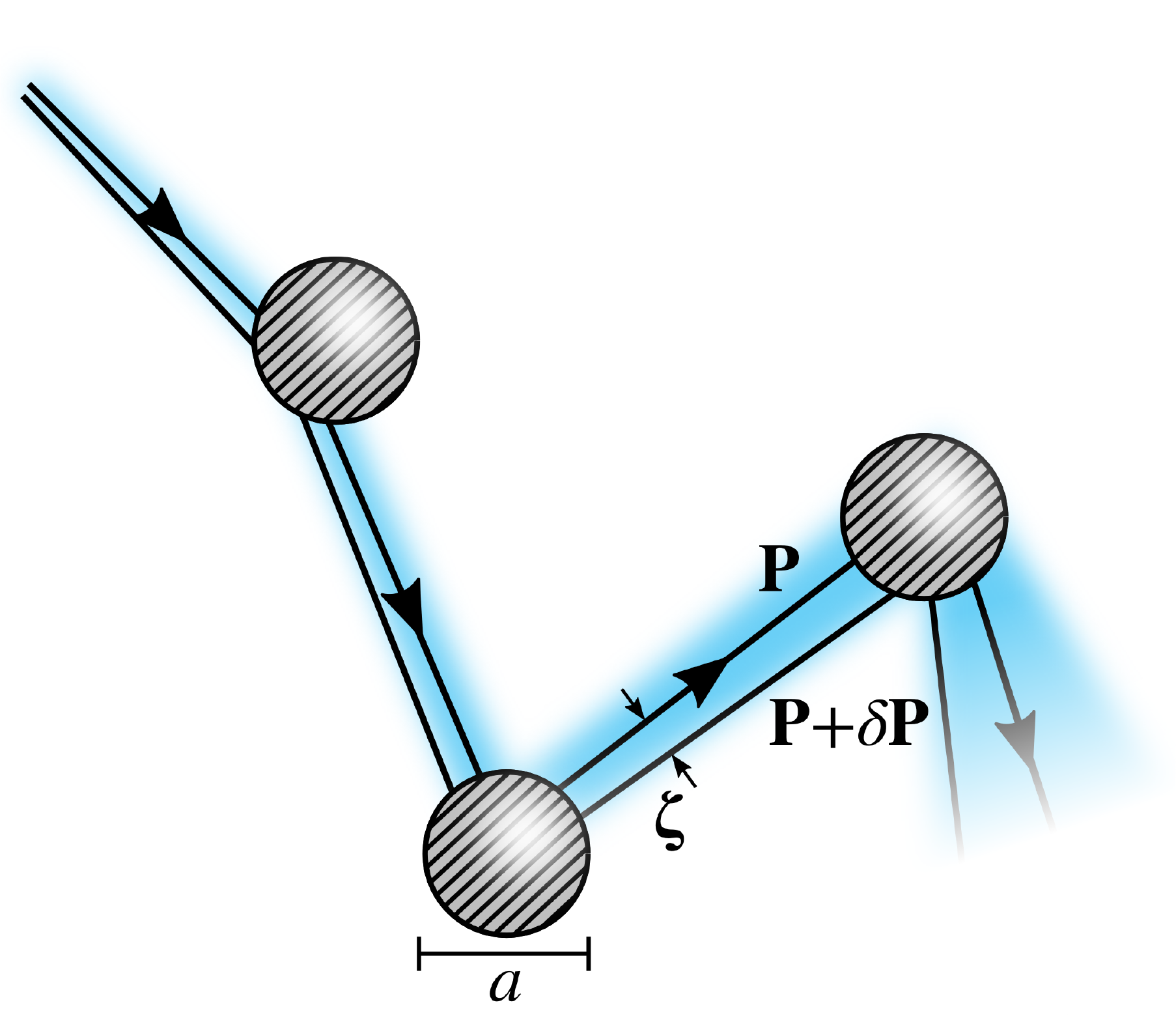}
	\caption{\label{fig:Diverging-trajectories}
		(Colour online)
		Two classical trajectories of electrons scattered off a chain of impurities. 
		The initially small momentum difference $ \delta \bP $ and spatial separation $ \bm{\zeta} $ grow exponentially because of
		the chaotic dynamics until the time of evolution reaches the Ehrenfest time $ t_E $, at which one of the trajectories misses the next consecutive impurity of the chain. 
		 }
\end{figure}

If the position and the momentum of the particle were known exactly at time $t=0$,
the particle would move along a classical trajectory, hereinafter referred to as the ``master trajectory'',
determined by the classical equations of motion and the initial conditions 
$(\bR,\bP)_{t=0}=(\bR_0,\bP_0)$. 
In this case, the momentum and position of the particle $(\bP, \bR )_t$ are deterministic functions.
However, if the initial momentum is uncertain, possible trajectories of the particle deviate from the master trajectory,
and the uncertainty of the momentum $\bP(t)$ at later times $t>0$ grows as shown in Fig. \ref{fig:Diverging-trajectories}.
Because of this growth of the uncertainty, the amount of information stored in the momentum
of the particle decreases with time.

Introducing small deviations $\delta \bP (t)$ and $\bzeta(t)$ of momentum and position
from the master trajectory at time $t$, the evolution of the uncertainties of momentum and coordinate
is described by the equations
\begin{subequations}
	\label{eq:setDiffEq}
\begin{align}
\tfrac{d}{dt}\langle \delta P^{2}\left(t\right)\rangle_{\prep} & =\frac{{4p}^{2}\lambda_0^{3}}{v^{2}}\langle\zeta^{2}\left(t\right)\rangle_{\prep}
\label{ClassSyst1}
\\
\tfrac{d}{dt}\langle\zeta^{2}\left(t\right)\rangle_{\prep} & =\frac{2v}{p}\langle\bm{\zeta}\cdot \delta \bP \left(t\right)\rangle_{\prep}
\\
\tfrac{d}{dt}\langle\bm{\zeta}\cdot \delta \bP \left(t\right)\rangle_{\prep} & =\frac{v}{p}\langle \delta P^{2}\left(t\right)\rangle_{\prep},
\label{ClassSyst2}
\end{align}
\end{subequations}
where $\lambda_0$ is the Lyapunov exponent for the system under consideration
and $v$ and $p$ are the absolute values of, respectively, 
the velocity and momentum of the particle, which remain constant during elastic collisions with the impurities
or the walls of the billiard. 

The system of equations (\ref{ClassSyst1})-(\ref{ClassSyst2}) describes also the evolution of the divergence between trajectories in 
momentum and coordinate spaces for a particle scattered elastically off randomly located impurities~\cite{Syzranov:ChaosTransition,LarkinOvchnnikov}. The derivations of the respective equations in Refs.~\cite{Syzranov:ChaosTransition} and \cite{LarkinOvchnnikov} relied on the procedure of disorder averaging, i.e.
averaging over the locations of impurities. We emphasise, however, that 
we consider here a particular realisation of disorder (or of the walls of a chaotic billiard)
and assume that the evolution of the separation between trajectories may be considered {\it self-averaged}.
Namely, we assume that the evolution of the uncertainty $\langle \delta P^{2}\left(t\right)\rangle_{\prep}$
of the particle's momentum matches the evolution of a beam of particles with the same width (in momentum and 
coordinate spaces) averaged over disorder:
\begin{align}
	\langle \delta P^{2}\left(t\right)\rangle_{\prep}=\langle \delta P^{2}\left(t\right)\rangle_{\dis}.
\end{align}
This assumption is justified if 
a considerable change of the deviation of momentum
requires multiple scattering events.

Solving equations (\ref{ClassSyst1})-(\ref{ClassSyst2}) gives exponential growth, $\langle \delta P^{2}\left(t\right)\rangle_\prep \propto e^{2\lambda_0t}$, of the variance of the particle's momentum
with the Lyapunov exponent $\lambda_0$. So long as a significant change of the variance 
requires a large number of collisions with impurities or with the walls of the system, 
the distribution
of the projection of momentum on the $z$ axis may be assumed Gaussian,
\begin{align}
\label{NormalDitrSingeParticle} 
\rho(P_z;t) 
& = 
\\
& \left[2\pi \langle \delta P_z^{2}\left(t\right)\rangle_{\prep} \right]^{-{1}/{2}}
\exp \left[\frac{(P_z -\langle P_z \left(t\right)\rangle_{\prep} )^2}{2\langle \delta P_z^{2}\left(t\right)\rangle_{\prep}} \right],
\notag 
\end{align}
according to the central limit theorem. The distribution of momentum is also Gaussian in a more generic case of an arbitrary
number of collisions on time $t$, provided the initial distribution at $t=0$ is Gaussian.
The variance $\langle \delta P_z^{2}\left(t\right)\rangle_{\prep}$ of the projection of momentum $P_z$
on a particular axis $z$ in Eq.~\eqref{NormalDitrSingeParticle}
is proportional to the variance $\langle \delta P^{2}\left(t\right)\rangle_\prep$ of the total momentum
and thus also grows exponentially with the exponent $2\lambda_0$.

Using the definition~\eqref{eq:def-information} of the amount of information encoded in the projection
$P_z$ of momentum and the time dependence $\langle \delta P_z^{2}\left(t\right)\rangle_\prep
\propto e^{2\lambda t_0}$ of the projection, we obtain, thus, that the information
\begin{equation}
I_{P_z}\left(t\right)=\text{const}-\lambda_0 t
\label{eq:info-gauss}
\end{equation}
changes with the rate given by the Lyapunov exponent $\lambda_0$.
In what follows, we generalise this connection between the rate of change of information, encoded in
the system's observables, and the Lyapunov exponents to a disordered metal coupled to a dissipative bath
with short-range correlations.


\section{Model for a many-body chaotic system\label{sec:Model}}

In what follows, we derive the main result of this paper for the connection
between out-of-time-order correlators and information dynamics for a system of 
electrons in a disordered medium 
interacting with neutral excitations, which we label as ``phonons''. In principle, these bosons
may be represented by excitations of a different nature, e.g. magnons, or may mediate electron-electron
interactions, in which case the bosons correspond effectively to plasmons.
Thus, we expect the results to hold for a system of interacting particles, as well as for electrons
coupled to a bosonic bath. 
The Hamilton of the systems is given by
\begin{equation}
\label{eq:H}
\hat{H}=\hat{H}_{\text{el}}+\hat{H}_{\text{el}-\text{ph}}+\hat{H}_{\text{ph}},
\end{equation}
where
\begin{align}
\hat{H}_{\text{el}}= 
\int_{ \br}  \psi^{\dagger}\left( \br \right)
\left[\epsilon_{\hat{\mathbf{k}}} + U_{\text{imp}}\left(\br\right)\right] \psi\left(\br\right);
\label{eq:H-el}
\end{align}
$\psi(\br)$ and $\psi^{\dagger}(\br)$
are the creation and annihilation operators for an electron at coordinate $\br$;
$\hat{\mathbf{k}}=-i\frac{\partial}{\partial \br}$ is the momentum operator; $\epsilon_{\hat{\mathbf{k}}}$
is the operator of the electron dispersion (measured from the chemical potential)
\begin{equation}
\hat{H}_{\text{el}-\text{ph}}=g\int_{\br}\phi\left(\br\right)\psi^{\dagger}\left(\br\right)\psi\left(\br\right)
\label{eq:ephHam}
\end{equation}
is the Hamiltonian of the electron-phonon coupling; $H_{\text{ph}}$ is the Hamiltonian of the
phonon bath with $g$ being a small constant of electron-phonon coupling. We do not consider the spin degree of
freedom as it does not affect the results for the Lyapunov exponents and the rates of information
change in the system; $ \phi(x) = \phi^\dagger(x)$ is the phonon field operator 
corresponding to a real-valued displacement field. 

The Hamiltonian $\hat{H}_\ph$ in Eq.~\eqref{eq:H} governs the dynamics of the phonon bath.
The exact microscopic details of the Hamiltonian of phonons are not important for the results
of this paper.   
The phonons are assumed to be correlated on a length $\xi_\ph$
significantly shorter than the characteristic size $a$ of the impurities or, in the absence of impurities, the size of the chaotic billiard. We assume also that the distribution function of the phonons
is affected weakly by the coupling to the electrons.

Quenched disorder is represented by the potential $U_{\text{imp}}\left(\br\right)=\sum_{i}U\left(\br-\mathbf{R}_{i}\right)$ of randomly located impurities,
with $ U(\br-\mathbf{R}_{i}) $ 
being the potential of the $i$th impurity. 
The potential of one impurity is assumed to be short-ranged compared to the mean free path of the electrons. 


\section{Out-of-time-order correlator\label{sec:Out-of-time-order-correlator}}

A frequently used definition of chaotic behaviour in a system, which we also employ here,
is the exponential growth of an OTOC~\eqref{eq:otoc} of two operators.
In this paper, we focus on the OTOC
\begin{align}
	F\left(t\right)=-\left<\big[\hat{P}_{z}\left(t\right),\hat{P}_{z}\left(0\right)\big]^{2}\right>, 
	\label{MomentumOTOC}
\end{align}
of the
operators
\begin{equation}
\hat{P}_{z}\left(t\right)=\int_{\br}\psi^{\dagger}\left(\br,t\right)\hat{p}_{z}\psi\left(\br,t\right)
\end{equation}
of the projection of the total momentum of the electrons in the system,
where $\langle\dots\rangle=\mathcal{Z}^{-1}\text{tr}[e^{-\hat{H}/T}\dots]$;
$\hat{p}_{z}=-i\nabla_{z}$
is the single-particle momentum operator and 
$z$ is one of the coordinates of $\br$.
OTOCs are often computed by means of a diagrammatic technique on a four-branch Keldysh contour
(see, e.g., Refs.~ \cite{Stanford:firstOTOC,PatelSachdev,PatelChowdhurrySachdevSwingle,KlugSchmalian,WermanKivelsonBerg,LiaoGalitski}).
Here, however, we use a different approach, based on deriving the kinetic equation for OTOCs,
similarly to the  derivation in Ref.~\cite{KlugSchmalian}.
As discussed in the previous section, we consider a particular realisation of disorder to determine the time evolution of the OTOC.


\subsection{Kinetic equation for OTOC}

For deriving the kinetic equation, it is convenient to introduce the correlation
function
\begin{align}
\label{eq:def-k}
K(\mathbf{R} & \br\tau,\mathbf{R}'\br'\tau',t)=\\
\langle[ & \psi^{\dagger}(\mathbf{R}\!-\!\tfrac{\br}{2},t\!-\!\tfrac{\tau}{2})\psi(\mathbf{R}\!+\!\tfrac{\br}{2},t\!+\!\tfrac{\tau}{2}),\hat{P}_{z}\left(0\right)]\nonumber \\
 & \times[\psi^{\dagger}(\mathbf{R}'\!-\!\tfrac{\br'}{2},t\!-\!\tfrac{\tau'}{2})\psi(\mathbf{R}'\!+\!\tfrac{\br'}{2},t\!+\!\tfrac{\tau'}{2}),\hat{P}_{z}\left(0\right)]\rangle\nonumber 
\end{align}
where $\langle\dots\rangle=\mathcal{Z}^{-1}\text{tr}[e^{-\hat{H}/T}\dots]$
represents averaging with respect to the equilibrium state
with the temperature $T$; $ \mathbf{R}^{(\prime)} $ and $ t^{(\prime)} $ are
the so-called ``centre-of-mass'' coordinates and times~\cite{RammerSmith:review,Kamenev:book}
of the respective pairs of the fermionic operators; $ \br^{(\prime)} $ and $ \tau^{(\prime)} $ 
are the respective differences of the coordinates and times.
The operators in the correlator \eqref{eq:def-k}
are out-of-time ordered, i.e. cannot be ordered on the Keldysh contour~\cite{Kamenev:book}.

Any OTOC of the form $\langle [\hA(t),\hat{P}_{z}(0)]^2 \rangle$ may be expressed conveniently 
as a linear combination of correlators~\eqref{eq:def-k} or their Wigner-transforms
\begin{align}
\label{eq:Kcl}
K & \left(\mathbf{R}\mathbf{p},\mathbf{R}'\mathbf{p}',t\right)=\\
& \int_{\br\omega\tau}\int_{\br'\omega'\tau'}e^{i(\omega\tau-\br\mathbf{p})+i(\omega'\tau'-\br'\mathbf{p}')}K\left(\mathbf{R}\br\tau,\mathbf{R}'\br'\tau',t\right),\nonumber 
\end{align}
where $\int_{\br\omega\tau}\ldots=\int d\br d\tau\frac{d\omega}{2\pi}\ldots$. For example, the OTOC~\eqref{MomentumOTOC}
of the operators of momentum projections is given by
\begin{equation}
F\left(t\right)=\int_{\mathbf{pR}}\int_{\mathbf{p'R'}}p_{z}p_{z'}K\left(\mathbf{R}\mathbf{p},\mathbf{R}'\mathbf{p}',t\right),
\label{K-WT-feg}
\end{equation}
where our convention for momentum integration is $\int_{\mathbf{p}}\ldots=\int\frac{d\bp}{(2\pi)^3}\ldots$. 

The function $K\left(\mathbf{R}\mathbf{p},\mathbf{R}'\mathbf{p}',t\right)$ evolves similarly to the joint distribution 
of two systems of qausiparticles or the product 
$f(\bp_1,\bR_{1})f(\bp_2,\bR_{2})$
of quasiparticle distribution functions. 
For non-interacting particles 
in the absence of the phonon bath, the evolution of these two objects is governed by exactly the 
same kinetic equation. This reflects the mapping between the evolution of 
an out-of-time-order correlator and the joint distribution of two copies of the system, as discussed in Ref. \cite{Syzranov:OTOCdot}. 

When the system is weakly coupled to a phonon bath, the evolution
of the correlator \eqref{eq:Kcl}
 is slightly modified by the phonons
and is governed by the equation
(see Appendix \ref{sec:Derivation-kinetic-equation} for the details of the derivation)
\begin{equation}
(\partial_{t}+ i\hat{L}_{\bR,\bp} + i\hat{L}_{\bR',\bp'} )K (\bR\bp,\bR'\bp',t) 
=I_{\text{ph}}\left[K\right],
\label{eq:kin-eq}
\end{equation}
where the left-hand side with the Liouville operator 
\begin{equation}
\label{eq:Liouvillean}
i\hat{L}_{\bR,\bp} = \bv_\bp \! \cdot \! \nabla_\bR - \nabla_\bR U_{\imp}(\bR)\! \cdot \! \nabla_\bp  
\end{equation}
describes ``free'' propagation of quasiparticles in the potential $U_{\imp}$ created by impurities, 
and the ``collision integral'' integral $I_{\text{ph}}\left[K\right]$ accounts for the effects of coupling to the phonon bath;
$\mathbf{v}_{\mathbf{p}}\equiv\nabla_{\mathbf{p}} \epsilon(\mathbf{p})\approx v_{F}{\mathbf{p}}/{|\mathbf{p}|}$
is the velocity of the quasiparticles.  

Equation~\eqref{eq:kin-eq} describes the dynamics of the correlator $K (\mathbf{R}\mathbf{p},\mathbf{R}'\mathbf{p}',t)$
in the quasiclassical regime, i.e. when the impurity sizes and the characteristic length scale of the correlator $K (\mathbf{R}\mathbf{p},\mathbf{R}'\mathbf{p}',t)$
exceed considerably the wavelength $\lambda_F$ of the quasiparticles at the Fermi surface. Under these conditions, the correlator $K (\mathbf{R}\mathbf{p},\mathbf{R}'\mathbf{p}',t)$
is a smooth function of its arguments.

\begin{figure}
	\centering
	\includegraphics[width=0.9\linewidth]{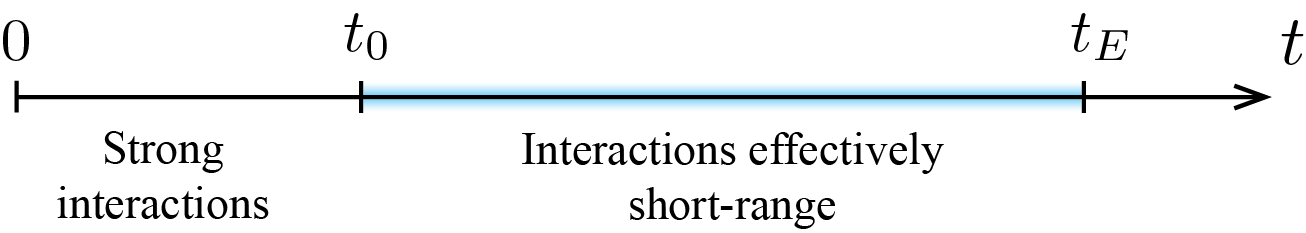}
	\caption{
		\label{fig:regimes}
		(Colour online)
		Characteristic time scales of the evolution of the system.
		At short times, $t \ll t_{\ph}$, the characteristic correlation length $\xi_\ph$ of the phonon bath exceeds
		the distances between the phase trajectories of quasiparticles; in this regime, phonons lead to strong correlations between the trajectories.
		At longer times, $t_{\ph} \ll t \ll t_E$, phonons affect the quasiparticle motion but do not cause significant correlations between different phase trajectories which 
		contribute to the OTOC.
		}
\end{figure}

Initially, i.e. at time $t=0$,
the correlator is sharply peaked as a function of $|\bR-\bR^\prime|$ and $|\bp-\bp^\prime|$, with the characteristic
length on the order of $\lambda_{F}$ \cite{Syzranov:ChaosTransition}. During the evolution, the distance between 
the trajectories of quasiparticles in phase space increases and the characteristic length and momentum 
scales of the correlator
$K (\mathbf{R}\mathbf{p},\mathbf{R}'\mathbf{p}',t)$ grow.

 In most of this paper, we focus on short-range interactions
or short-range-correlated phonon bath, whose length scale $\xi_\ph$ is exceeded considerably by the
characteristic length scale $a$ of the system. This latter scale may be given, for example, by the impurity size in a disordered metal
or by the size of the system, in the case of a classically-chaotic billiard. 
The regime of evolution, for which the characteristic length scale of the correlator $K (\mathbf{R}\mathbf{p},\mathbf{R}'\mathbf{p}',t)$
is larger than $\xi_\ph$ but smaller than $a$ corresponds to a 
parametrically large interval
of times, $t_{\ph}\ll t\ll t_E$, shown in Fig.~\ref{fig:regimes},
and the time $t_{\ph}$ given by Eq.~\eqref{eq:tph}.
As we demonstrate below, the phonons affect the dynamics at this interval,
 but do not cause significant correlations 
between different phase trajectories contributing to the OTOC.

Therefore, the collision integral in the kinetic equation~\eqref{eq:kin-eq} in the interval $t_{\ph}\ll t\ll t_E$
is given by (see Appendix~\ref{sec:Derivation-kinetic-equation}) 
\begin{align}
I_{\text{ph}} &\left[K\right]
=
-K(\bR\bp,\bR'\bp',t)
\int_{\mathbf{k}}\left(\Gamma_{\mathbf{p\rightarrow k}}
+\Gamma_{\mathbf{p'\rightarrow k}}\right)
\label{eq:collision-integral} \\
&+\int_{\mathbf{k}}\Gamma_{\mathbf{k\rightarrow p}}K (\bR\bk,\bR'\bp',t)+\int_{\mathbf{k}}\Gamma_{\mathbf{k\rightarrow p'}}K (\bR\bp,\bR'\bk,t), \notag
\end{align}
where we have introduced the phonon-assisted scattering rate 
\begin{multline}
\Gamma_{\mathbf{p\rightarrow k}}  =ig^{2}\big[D^{<}(\bp-\bk,\epsilon_\bp-\epsilon_\bk)f_{0}\left(\epsilon_\mathbf{k}\right)  \\ +D^{>}(\mathbf{p-k},\epsilon_\bp-\epsilon_\bk)(1-f_{0}(\epsilon_\mathbf{k}))\big],
\end{multline}
where $g$ is the electron-phonon coupling constant, defined by Eq.~\eqref{eq:ephHam};
$ D^< $ and $ D^> $ are the 'lesser' and 'greater'~\cite{RammerSmith:review,Kamenev:book} phonon propagators [see Appendix, Eq. (\ref{eq:propagators}) for the definitions]; $ f_0 $ if the Fermi distribution function. 

\subsubsection*{Initial conditions}

The kinetic equation \eqref{eq:kin-eq} with the collision integral \eqref{eq:collision-integral} describes
the evolution of OTOCs at times $t_\ph\ll t\ll t_E$, when the characteristic distances of the trajectories contributing
to the chaotic behaviour exceed, on the one hand, the correlation length $\xi_\ph$ of the phonon bath and 
are significantly smaller, on the other hand, than the characteristic length scale of the ballistic billiard.
To provide a full description of the evolution on this time interval,
the kinetic equation \eqref{eq:kin-eq} has to be complemented by the initial condition
for the correlator $K(\bR\bp,\bR'\bp',t)$
at $t=t_\ph$.   

It has been demonstrated in Ref.~\cite{Syzranov:ChaosTransition} that at time $t=0$ the correlator 
for a metal in equilibrium
is given by

\begin{align}
\label{eq:KinitCond}
K (\bR\bp,&\bR'\bp',0) 
 = (4\pi)^{3}\delta(\bp-\bp')\\
&\partial_Z \partial_Z^\prime
 \int_{\bq}e^{2i\bq(\bR-\bR')}f_{0}(\epsilon_{\bp-\bq})\left[1-f_0(\epsilon_{\bp+\bq})\right],
\notag
\end{align}
and, so long as distances $|\bR-\bR^\prime|$ exceeding the Fermi wavelength $\lambda_F$ are concerned,
may be approximated as derivatives of a delta function
with respect to the $z$-components of $\bR$ and $\bR^\prime$:

\begin{align}
	K (\bR\bp,\bR'\bp',0)\approx
	& (2\pi)^6
	f_0(\epsilon_\bp)\left[1-f_0(\epsilon_\bp)\right]
\nonumber	\\
	&\delta(\bp-\bp^\prime)\partial_Z \partial_{Z^\prime}
	\delta(\bR-\bR^\prime).
	\label{eq:InitCondSimplified}
\end{align}
In the case of short-range correlations of the bath, on which we focus in most of this paper, the initial period $t_\ph$ of the evolution
of the correlator (cf. Fig.~\ref{fig:regimes}) is short and the correlator may be assumed to have the form \eqref{eq:InitCondSimplified}
at $t=t_\ph$, which may then be used as the initial condition for the kinetic equation \eqref{eq:kin-eq}.
For insufficiently short-range correlations of the phonons, however, the initial conditions will be different.
The exact form of the initial condition for the correlator $K(\bR\bp,\bR'\bp',0)$ does not affect our results for the Lyapunov
exponents and the rates of information dynamics, so long as the correlator remains sharply peaked at short times $t\lesssim t_\ph$.


\subsection{Dynamics of the OTOCs}

{\it Conservation laws.} From the kinetic equation \eqref{eq:kin-eq} with the collision 
integral \eqref{eq:collision-integral} it follow that the correlator $K$ has the conservation law
\begin{align}
	\int_{\bR,\bR^\prime,\bp,\bp^\prime}
	K(\bR\bp,\bR'\bp',t)\equiv N_K=\text{const}.
	\label{eq:ConservLaw}
\end{align}
This conservation law is analogous to the conservation of the number of particles
$N=\int_{\bR,\bp}f(\bR,\bp,t)$ in an electronic
systems which obeys the conventional kinetic equation for the distribution function $f$~\cite{Abrikosov:metals}.
For the initial condition \eqref{eq:InitCondSimplified} the constant $N_K$ in the conservation law~\eqref{eq:ConservLaw}   
is zero, $N_K=0$.

{\it Smooth and peaked parts of the correlator.}
To describe further the evolution of the correlator $K (\bR\bp,\bR'\bp',t)$,
obeying the kinetic equation \eqref{eq:kin-eq},
it is convenient to decompose it into the ``peaked'' ($ p $)
and ``smooth''($ s $) parts,
\begin{equation}
\label{eq:KsAndKp}
K 
=
 K^{(s)} 
 + K^{(p)}, 
\end{equation}
where the peaked part $K^{(p)}$ has characteristic length $|\bR-\bR^\prime|$ and momentum
$|\bp-\bp^\prime|$ scales considerably shorter than, respectively,
\begin{subequations}
	\label{eq:chaosScales}
	\begin{align}
	a_\ch &= \frac{a^2}{\ell},
	\label{ach}
	\end{align}
	and
	\begin{align}
	p_\ch &= \frac{a}{\ell}p_F,
	\label{pch} 
	\end{align}	
\end{subequations}
while the smooth part $K^{(s)}$ does not vary significantly on these scales, where 
$\ell=v_F\tau_0$ is the mean free path, i.e. the distance that a quasiparticle travels between collisions with impurities
or the walls of the billiard in the absence of the phonon bath, and $p_F$ is the Fermi momentum.

\begin{figure}[t]
	\centering 
	\subfloat[\label{fig:Ka}]{\includegraphics[width=0.47\columnwidth]{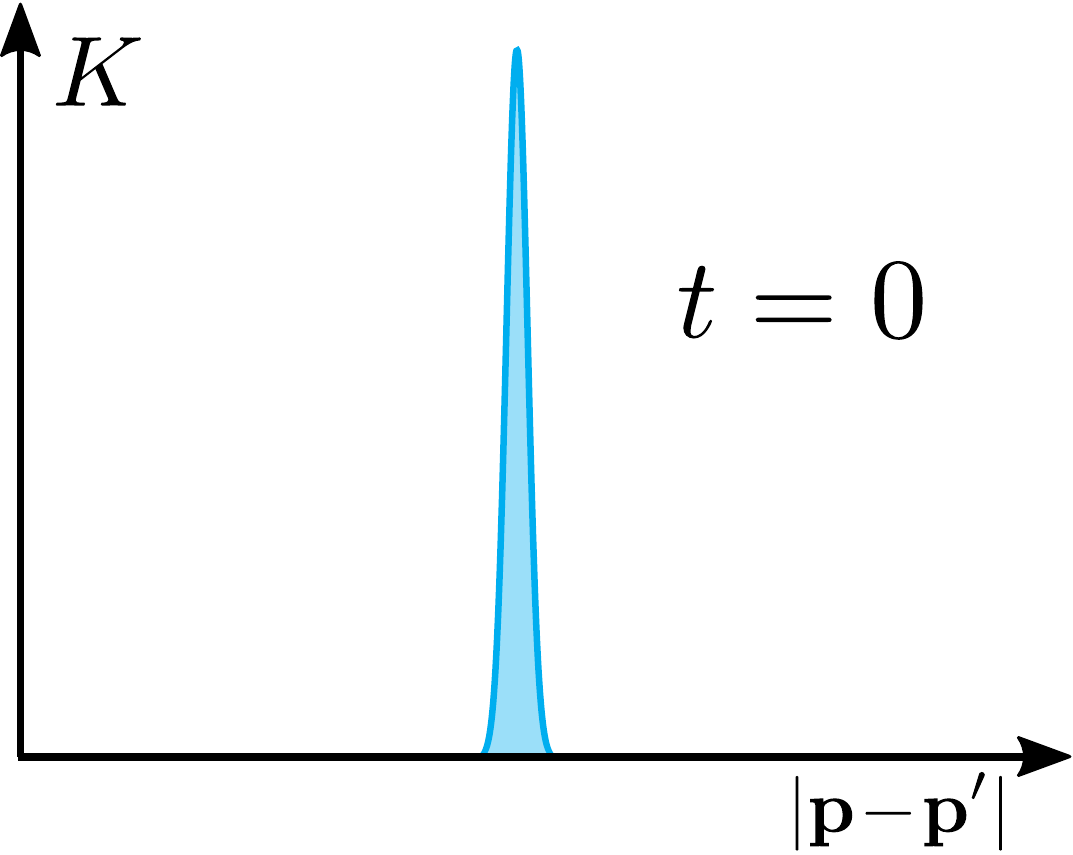}}
	\hfill
	\subfloat[\label{fig:Kb}]{\includegraphics[width=0.47\columnwidth]{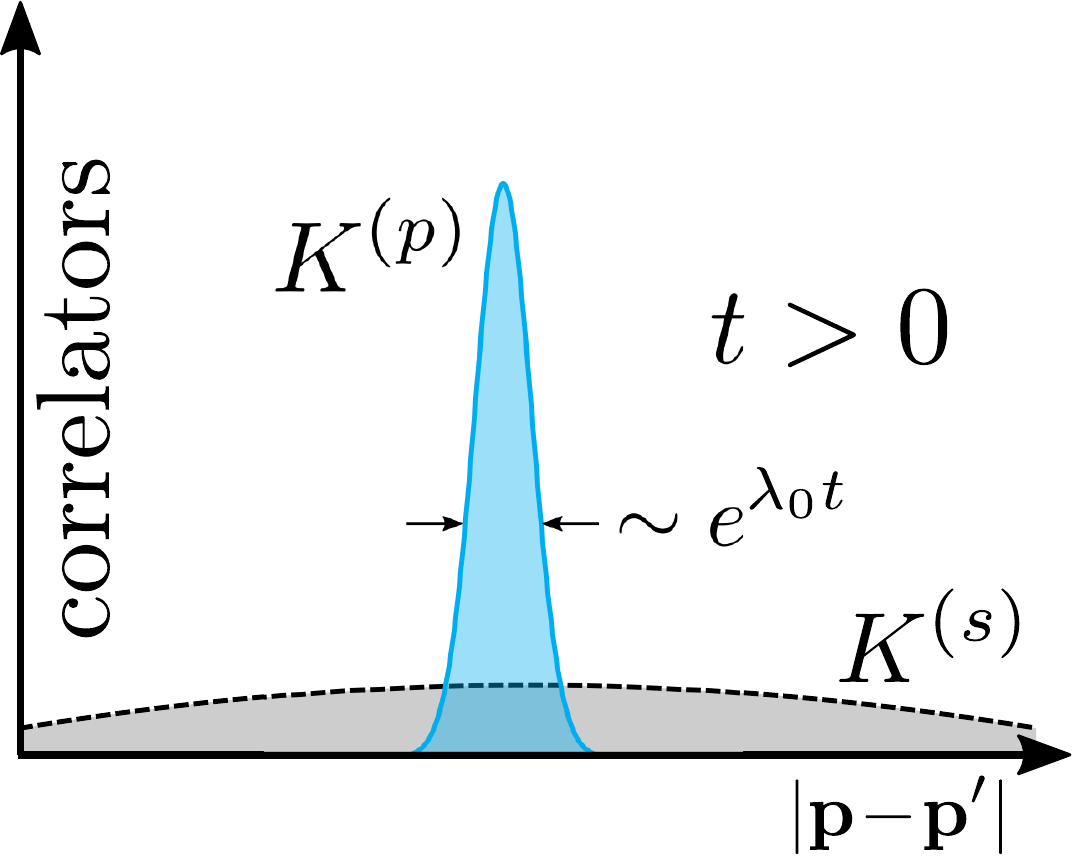}}
	\caption{ (Colour online)	
		The correlator $ K (\mathbf{R}\mathbf{p},\mathbf{R}'\mathbf{p}',t) $ as a function
		of momentum difference $\bp-\bp^\prime$ for close locations $\bR$
		and $\bR^\prime$.
		(a) The correlator at $t=0$. It is a sharply peaked function with the characteristic
		momentum scale $\lambda_F^{-1}$. 
		(b) The correlator at later times. It consists of two parts, a sharply peaked part $K^{(p)}$
		with an exponentially growing width and correlated on length and momentum
		scales significantly shorter, respectively, than \eqref{ach} and \eqref{pch}, and a smooth part $K^{(s)}$
		\label{fig:K} which does not vary significantly on these scales.}
\end{figure}

\begin{figure}[t]
	\centering 
	\includegraphics[width=0.95\columnwidth]{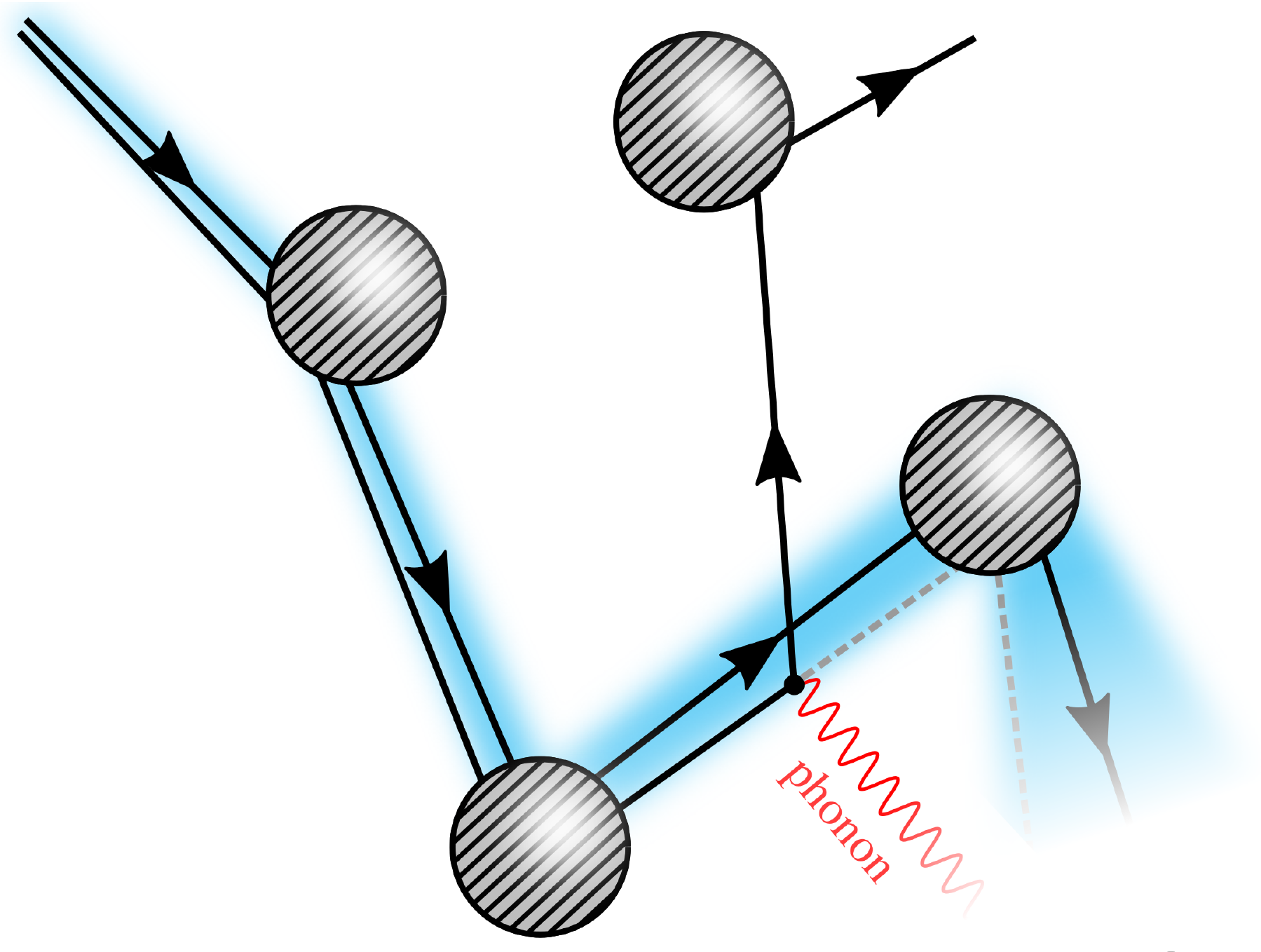}
	\caption{
		\label{fig:scatteringProc}
		(Colour online)
		Electron trajectories in the presence of both impurity and phonon scattering.
		Under impurity scattering, 
		two quasiparticles with close initial momenta and locations propagate along close trajectories,
		with an exponentially growing separation $\propto e^{\lambda_0 t}$ between them.
		If one of the quasiparticles is scattered by a phonon,
		its motions deviates significantly from the initial trajectory, the motion of the quasiparticles
		becomes effectively uncorrelated, and the separation between them grows diffusively ($\propto t^\frac{1}{2}$).}
\end{figure}

Initially, the correlator $K(\bR\bp,\bR'\bp',t)$, shown in Fig.~\ref{fig:Ka}, is sharply peaked at origin
as a function of $\bp-\bp^\prime$ and $\bR-\bR^\prime$, according to Eq.~\eqref{eq:InitCondSimplified}.
At later times, it acquires a ``smooth'' part as well, due to the scattering by long-wavelength
phonons and the concomitant deflections of the classical electron trajectories, as shown in Fig.~\ref{fig:scatteringProc}.

Smooth $K^{(s)}$ and peaked $K^{(p)}$
 correlators evolve qualitatively differently under impurity scattering.
When electrons are scattered off impurities (or the walls of the billiard), the peaked part remains peaked on sufficiently
short times,
but its width grows exponentially with time $\propto e^{\lambda_0 t}$, as shown in Fig.~\ref{fig:Kb}.
The smooth part also remains smooth and on times $t\gg \tau_0$ has the characteristic momentum scale $|\bp-\bp^\prime|$
on the order of the Fermi momentum $p_F$.

Indeed, classical electron trajectories corresponding to the same momenta
and a small length separation $\zeta\ll a_\ch$ remain close to each other
over several collisions and hit the same chain of impurities, as shown in Fig.~\ref{fig:Diverging-trajectories}.
If they 
 scatter off an impurity of size $a$, then after the collision
their momenta are still close and confined to a small angle $\phi\sim \zeta/a\ll a/\ell$.
If, after that, these trajectories collide with another impurity, distance $\ell$ away from the first collision,
the spatial separation
between the trajectories is on the order of $\phi \ell\ll a$ and is still significantly smaller than the impurity size $a$.
Similarly, close electron trajectories with momenta differences smaller than the scale $p_\ch$, given by Eq.~\eqref{pch},
remain closer than the impurity size after one or several collisions.
The divergence between the respective trajectories is described by the system of equations~(\ref{ClassSyst1})-(\ref{ClassSyst2})
in the absence of the phonon bath. It leads to the exponential growth of the separation between classical trajectories,
which leads to the exponential growth of the width of the correlator $K^{(p)}$, shown in Fig.~\ref{fig:Kb}, in the absence of the phonon bath.
As we demonstrate below, this exponential growth persists in the presence of collisions with phonons;
electron-phonon interaction reduce, however, the height of the peak.

If, on the other hand, the spatial or momentum separation between two quasiparticles exceeds the scales \eqref{ach} or \eqref{pch},
then these quasiparticles start to move far away from each other after one collision with an impurity;
even if these quasiparticles hit the same impurity, the separation
between their trajectories will exceed the impurity size $a$ before the next elastic collision.
Such quasiparticles move diffusively relative to each other, and the spatial separation between them
depends on time as $\propto t^\frac{1}{2}$ for $t\gg\tau_0$.

While peaked and smooth correlators remain, respectively, peaked and smooth under collisions
with impurities, short-range-correlated phonons change quasiparticle momenta
by values exceeding the scale (\ref{pch}) and thus may convert the peaked and the smooth correlators to each
other.

{\it Contribution of the smooth part to the OTOC.}
Initially, the correlator $K(\bR\bp,\bR'\bp',t)$ is sharply peaked, $K=K^{(p)}$. The scattering of momenta
$\bp$ and $\bp^\prime$ by phonons may lead to large momentum differences $|\bp-\bp^\prime|$
and, thus, result in the emergence of the smooth part $K^{(s)}$ of the correlator.
If correlations in the phonon bath are sufficiently short-range,  i.e.
if the phonon momenta are sufficiently long, the momenta $\bp$ and $\bp^\prime$ of the smooth 
part of the smooth correlator $K^{(s)}(\bR\bp,\bR'\bp',t)$ are distributed uniformly on the Fermi
sphere, and this part does not contribute to the OTOC, 
\begin{align}
	F^{(s)}=\int_{\bR,\bR^\prime,\bp,\bp^\prime} p_z p'_z K^{(s)}(\bR\bp,\bR'\bp',t)= 0.
\end{align}

For a more generic phonon bath, which is correlated on momentum scales shorter than the Fermi momentum $p_F$
but longer than the scale \eqref{pch}, the contribution $F^{(s)}$ of the smooth part $K^{(s)}$ of the correlator $K$
is still strongly suppressed, because it comes from the projections of two momenta $p_z$ and $p_z^\prime$
for different quasiparticle trajectories, separated by a large distance in a disordered system, who have little correlation between
each other.
Therefore, the main contribution to the OTOC~\eqref{MomentumOTOC} comes from the peaked part $K^{(p)}$ of the correlator $K$.

{\it Contribution of the peaked part to the OTOC.}
In this paper, we focus on a phonon bath with sufficiently short-range spatial correlations, which lead to the characteristic
momenta of the phonons exceeding the scale $p_\ch$ given by \eqref{pch}.
This phonon bath, when interacting with the quasiparticles, converts the sharply peaked part $K^{(p)}$ of the correlator $K$
to the smooth part $K^{(s)}$.
From the kinetic equation \eqref{eq:kin-eq} with the collision integral \eqref{eq:collision-integral} it follows that
at sufficiently short times, when the smooth 
part of the correlator is small, the evolution of the peaked part is described by
the equation
\begin{align}
\left(\partial_{t}+ i\hat{L}_{\bR,\bp} + i\hat{L}_{\bR',\bp'}  
 -\frac{2}{\tau}
 \right)K^{(p)} (\mathbf{R}\mathbf{p},\mathbf{R}'\mathbf{p}',t) 
=0,
\label{KineticEquationWithPhonons}
\end{align}
where the rate $1/\tau$ is given by  
\begin{multline}
\label{eq:phonon-scattering-rate}
\tau^{-1} = 
\int_\bk \Gamma_{\bp\rightarrow\bk}\big|_{|\bp| = p_F} = 
-2 g^{2}\int_{\bk}  
\im D^R_{\mathbf{p}-\bk} (\epsilon_{\mathbf{p}}-\epsilon_{\mathbf{q}}) \\
\times \left[n_{B}(\epsilon_{\mathbf{p}}-\epsilon_{\bk})+1-f_{0}(\epsilon_{\bk})\right]\big|_{|\bp| = p_F},
\end{multline}
and matches the rate of scattering of electrons by phonons in a metal~\cite{Mahan:manyParticlePhysics}, 
where $f_{0}(\omega)=\left[\exp(\omega/T)+1\right]^{-1}$ and $n_{B}(\omega)=\left[\exp(\omega/T)-1\right]^{-1}$ are, respectively,
the equilibrium Fermi and Bose distribution functions and we have taken into account that the momenta
$\bp$ and $\bp^\prime$ are close to the Fermi surface.

The retarded phonon propagator $D^R(\br-\br^\prime) = -i\theta(t-t^\prime) \langle \phi(\br,t)\phi(\br',t') \rangle $, where $\langle\dots\rangle=\mathcal{Z}_{\text{ph}}^{-1}\text{tr} [e^{-\hat{H}_{\text{ph}}/T}\dots ]$, in Eq.~\eqref{eq:phonon-scattering-rate} is renormalised by phonon-phonon interactions and, thus,
has in general a smooth imaginary part as a function of both momentum and energy.
We emphasise that the term ``phonon'' is used loosely in this paper as a convention for a generic
bosonic excitation. The respective bosonic bath may be represented by magnons, plasmons, spinons etc.
In the case of instantaneous interactions, e.g., Coulomb interactions between the quasiparticles,
the unperturbed propagator $D^R_{\bq}$ is real, but acquires a finite imaginary part as a result 
of renormalisation (screening).

According to Eq.~\eqref{KineticEquationWithPhonons}, the evolution of the peaked
correlator $K^{(p)}$ at sufficiently short times is given by 
\begin{equation}
	K^{(p)} (\mathbf{R}\mathbf{p},\mathbf{R}'\mathbf{p}',t) = e^{-2t/\tau } K^{(p)}_0(\mathbf{R}\mathbf{p},\mathbf{R}'\mathbf{p}',t), 
\end{equation}
where $K^{(p)}_0$ is the correlator \eqref{eq:Kcl} for the same system in the absence of the 
phonon bath. The evolution of the latter correlator is governed by the equation
\begin{equation}
(\partial_{t}+ i\hat{L}_{\bR,\bp} + i\hat{L}_{\bR',\bp'} )K^{(p)}_0 (\mathbf{R}\mathbf{p},\mathbf{R}'\mathbf{p}',t) 
=0, 
\end{equation}
which matches the kinetic equation \eqref{eq:kin-eq} in the absence of the electron-phonon collision integral \eqref{eq:collision-integral}.

{\it Evolution of the OTOC under the influence of the phonon bath.}
We have demonstrated that the influence of the phonon bath on the evolution of the correlator
$K(\mathbf{R}\mathbf{p},\mathbf{R}'\mathbf{p}',t)$
 is reduced to the exponentially
decaying $\propto e^{-\frac{2t}{\tau}}$ prefactor, where $1/\tau$ is the electron-phonon scattering
rate, and to the emergence of the smooth part of the correlator, which does not contribute to the
OTOC \eqref{MomentumOTOC}.

Therefore, the time evolution of the OTOC for a system of electrons coupled to a
short-range-correlated phonon bath is 
described by the equation
\begin{equation}
F\left(t\right)=e^{-2t/\tau} F_0 (t)
\label{FtoSinglePartRelation}
\end{equation}
where 
\begin{equation}
F_0(t) =  \int_{\bR\bp\bR'\bp'} p_z p_z' K_0^{(p)} (\bR\bp,\bR'\bp',t)
\end{equation}
is the OTOC in the absence of the coupling to the bath, studied in detail in Ref.~\cite{Syzranov:ChaosTransition}. The Lyapunov exponent for an electronic system coupled to a 
short-range-correlated phonon bath is thus given by
\begin{align}
	\lambda=\lambda_0-{1}/{\tau}.
	\label{LambdaGen}
\end{align}

The evolution of the out-of-time-order correlator, described in this section, has a simple qualitative interpretation.
The growth of the OTOC comes from correlations between close classical trajectories
of quasiparticles.
If an electron gets scattered by a phonon, its trajectory deviates strongly from the classical trajectories of the other electrons
and becomes uncorrelated with them and, thus, ceases to contribute to the growth of the OTOC.
The probability that an electron avoids scattering by phonons on time $t$ is given by $e^{-\frac{t}{\tau}}$.
The growth of the OTOC, which is determined by correlations between pairs of such trajectories, gets multiplied by the 
factor $e^{-\frac{2t}{\tau}}$, compared to the phonon-free evolution. As a result, the time dependence of the OTOC
is given by $\propto e^{2\lambda_{0} t-\frac{2t}{\tau}}$   

When the coupling to the phonon bath is weak, the OTOC is almost unaffected by the phonon bath, in accordance with Eq.~\eqref{LambdaGen},
and grows with the exponent $\lambda_{0}$.
When the coupling to the phonons is strong, all correlations are suppressed, and the OTOC decays exponentially. At $\lambda_{0}=1/\tau$,
a transition between chaotic and non-chaotic behaviour takes place, analogous to the interaction-driven transition predicted 
in Ref.~\cite{Syzranov:ChaosTransition}.
This transition may be triggered, for example, by changing temperature, which affects the electron-phonon scattering rate $1/\tau$.


\section{The dynamics of information\label{sec:Information}}

\begin{figure}[b]
	\label{fig:Weak-measurment-scheme}
	\centering{}\includegraphics[width=0.9\columnwidth]{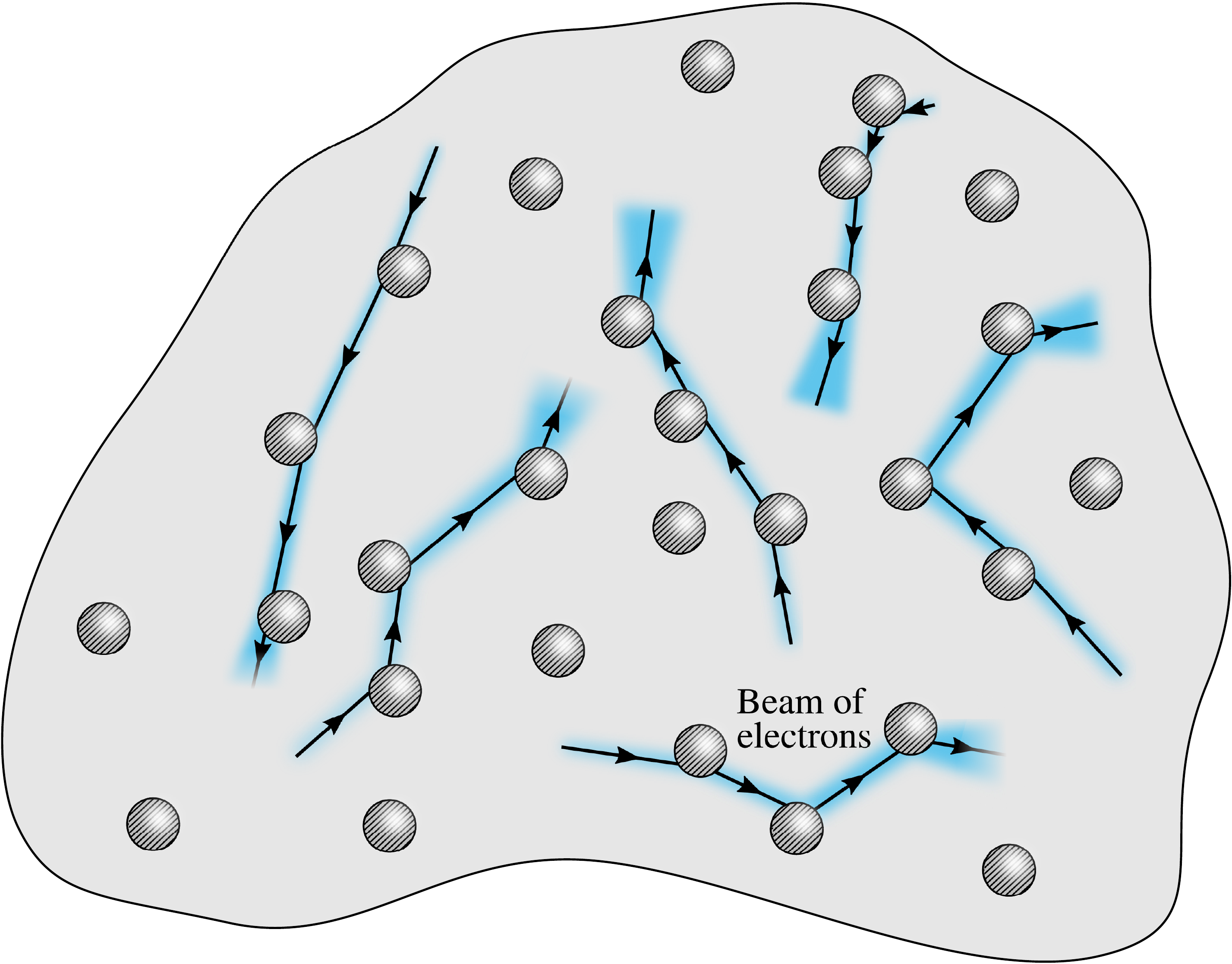}
	\caption{
		(Colour online) Propagation of a short-range correlated perturbation of the distribution function in a disordered metal.
		The perturbation may be decomposed into multiple independent local perturbations, which we call ``beams'' and which are shown
		in blue (dark grey) colour.
		Each of the beams gives an exponential contribution to the uncertainty 
		of the projection of the total momentum.
		\label{Fig:beams}
	 }
\end{figure}

In what follows, we discuss the relation between chaotic dynamics, characterised by the Lyapunov exponent
$ \lambda $ in a disordered metal coupled to a bosonic bath and the dynamics of information associated with a physical observable in this system. For the observable, we choose the same operator 
$ \hat{A} = \hat{B} = \hat{P}_z $ which enters the OTOC~\eqref{MomentumOTOC} growing
with the exponent $\lambda$.

{\it Measurement of the information in the observable.} We assume that at $t=0$ the systems is exposed to a perturbation, 
which modifies slightly the electron distribution function $f(\bR,\bp)$.
This perturbation has a small classical uncertainty, e.g. due to an imperfect procedure of perturbing the system or due to the initial perturbation not being known exactly. After time $t$,
the expectation value of the projection of the total momentum of the electrons, 
\begin{align}
	P_z(t)\equiv\langle \hP_z(t)\rangle,
\end{align}
is measured, where $\langle \ldots\rangle$ is averaging with respect to the initial (uncertain) state of the system. As the initial state is random, so is the expectation value of the projection
of the total momentum. 
The uncertainty of the averaged momentum may be characterised by the variance
\begin{align}
	\Average{\delta P_z(t)^2}_\prep
	=\Average{P_z(t)^2}_\prep-\Average{P_z(t)}_\prep^2. 
	\label{MomentumVariance}
\end{align}
In what follows, we examine the amount of information associated with the uncertainty of the measured observable.

The perturbation, created in the system, is described by the deviation
\begin{equation}
g(\bR,\bp,t)=f(\bR,\bp,t)-f_{0}(\epsilon_\bp), 
\end{equation}
 of the electron distribution function $f$
from the equilibrium distribution function $ f_0 $. 
As the system has a zero momentum in equilibrium, the momentum at time $t$ may be expressed 
through the perturbation $g$ as
\begin{align}
	P_z(t) = \int_{\bp\bR} p_z g(\bR,\bp,t).
\end{align} 

Clearly, if the perturbation is very smooth as a function of $\bp$ and $\bR$, it will spread in the system
diffusively, with the diffusion coefficient modified by the weak-localisation and interaction corrections, and will bear 
little or no signatures of chaotic behaviour. 
Therefore, we assume that the perturbation is short-range correlated in phase space
at the initial time $t=0$:
\begin{multline}
\label{eq:dist-cond} 
\int_{\mathbf{pR}}g(\mathbf{R},\mathbf{p},0)g(\mathbf{R}-\Delta \bR,\mathbf{p}-\Delta\bp,0) \\ 
=\mathcal{F} \left(\tfrac{|\Delta\mathbf{R}|}{l_\ex},\tfrac{|\Delta\mathbf{p}|}{q_\ex}\right),
\end{multline}
where  $\mathcal{F} (x,y)$ is a rapidly decaying function for the values of its arguments exceeding unity
and the scales $ l_\ex $ and $ q_\ex $, the characteristic ranges of correlations of the perturbation in
spatial and momentum spaces, are rather short. 

We assume that the correlation radii $ l_\ex $ and $ q_\ex $ in coordinate and momentum spaces are exceeded significantly
by the scales $ a_\ch $ and $ p_\ch $ given by Eqs.~(\ref{ach}) and (\ref{pch}). Under these conditions, the initial perturbation
may be considered as a superposition of perturbations, which are local in phase space and have sizes smaller than the scales 
$ a_\ch $ and $ p_\ch $. Each such local perturbation, which we call ``a beam'', propagates in the system independently of the other local perturbations, as shown in Fig.~\ref{Fig:beams}.

Because the beams are narrow in both momentum and coordinate spaces, their propagation resembles that of individual quasiparticles.
So long as the width of a beam remains shorter than the scales $ a_\ch $ and $ p_\ch $, its propagation is chaotic.
As we show below, 
each beam gives an exponential contribution to the uncertainty of the expectation $\langle \hP_z(t)\rangle$ of momentum,
which determines the amount of information
encoded by the probability distribution of this expectation.
Hence, to compute the rate of change of information in the observable $\langle \hP_z(t)\rangle$ it is sufficient
to consider only one beam, i.e. a perturbation which is sharply peaked initially in momentum and coordinate spaces.

{\it Evolution of the deviation of the distribution from equilibrium.}
Initially, the deviation $g(\bR,\bp,t)$ of the distribution function is a sharply peaked function around 
a certain momentum and location. As the system evolves, this function gets broadened due to the collisions
of electrons with impurities and phonons. 
Similarly to the out-of-time-order correlators in Sec.~\ref{sec:Out-of-time-order-correlator}, it is 
convenient to split it into "smooth" (s) and "peaked" (p) parts, 
\begin{equation}
g = g^{(s)} + g^{(p)}, 
\end{equation}
whose widths in momentum and coordinates spaces are, respectively, significantly larger and significantly
smaller than the scales $ a_\ch $ and $ p_\ch $ given by Eqs.~(\ref{ach}) and (\ref{pch}).

In the course of evolution, long-wavelength phonons change electron momenta significantly and, thus,
shift the weight of the peaked part $g^{(p)}$
to the smooth part $g^{(s)}$.
By analogy with the kinetic equation \eqref{KineticEquationWithPhonons} for the peaked part of the
out-of-time-order correlator $K(\bR\bp,\bR^\prime\bp^\prime,t)$,
the evolution of the peaked part of the distribution function is described by the kinetic equation 
\begin{eqnarray}
\label{eq:kinEqG}
\left(\partial _t + i\hat{L}_{\bR,\bp} - \frac{1}{\tau} \right) g^{(p)} (\bR,\bp,t) = 0 
\end{eqnarray}
with the Liouville operator $i\hat{L}_{\bR,\bp} $ given by Eq.~(\ref{eq:Liouvillean}) and
$1/\tau$ being the inelastic scattering rate of electrons due to electron-phonon interactions.

We emphasise that in both kinetic equations \eqref{KineticEquationWithPhonons}
and \eqref{eq:kinEqG} for the four-point and two-point correlators the effect of the bath is 
reduced to a relaxation term with the relaxation rate independent of the collisions of the electrons
with impurities or the walls of the system. This separation between elastic and inelastic scattering rates
takes place in systems with short-range character of the interactions 
(known as the ``ballistic Altshuler-Aronov'' regime~\cite{ZalaNarozhnyAleiner,Gantmakher:book})
 or correlations in the bath, considered here, relative to the elastic mean free path.
In a generic systems with long-range interactions or a short elastic scattering time, the
effective inelastic scattering rate receives significant renormalisations from collisions
with impurities~\cite{AltshulerAronov,ZalaNarozhnyAleiner,PatelSachdev,LiaoGalitski,Gantmakher:book}.

Equation \eqref{eq:kinEqG} gives the time evolution of the peaked part of the perturbation in the form
\begin{equation}
\label{eq:g0}
g^{(p)}(\bR,\bp,t) = e^{-t/\tau }g_0^{(p)}(\bR,\bp,t), 
\end{equation}
where $g_0^{(p)}(\bR,\bp,t)$ describes the deviation of the distribution function from equilibrium in the absence 
of the phonon bath. The function $g_0^{(p)}(\bR,\bp,t)$ obeys the kinetic equation
\begin{equation}
\label{eq:redKineticEq}
(\partial_t + i\hat{L}_{\bR,\bp}   ) g_0^{(p)}(\bR,\bp,t) =0.
\end{equation}
Similarly to the case of out-of-time-order correlators, the perturbation of the distribution functions
factorises into that in the absence of the bath and an exponentially decaying factor which accounts for the influence of the bath.

The smooth part $g^{(s)}$ of the perturbation is uniform on the Fermi surface if the characteristic phonon wavelength is on the order
of or larger than the Fermi momentum $p_F$~\footnote{In the case of a phononic bath, this is realised at temperatures
	exceeding the Debye temperature~\cite{Abrikosov:metals}.}. In the case of a bath with shorter-wavelength phonons,
with momenta between the scale $p_\ch$ and $p_F$, the smooth part of the distribution relaxes to a uniform distribution in
momentum space on time scales exceeding the transport scattering time.

Under these assumptions, the contribution of the smooth part to the total momentum
and to the variance \eqref{MomentumVariance} may be neglected. Therefore, similarly to the case of correlators considered in 
Sec.~\ref{sec:Out-of-time-order-correlator}, it is sufficient to consider only the momentum of the peaked part of the perturbation
of the distribution function.

The beams, considered here, are narrow in both coordinate and momentum spaces
and remain so on sufficiently short times $t\lesssim \lambda_{0}^{-1}\ln\left[\min\left(\frac{a_\ch}{l_\ex},\frac{p_\ch}{q_\ex}\right)\right]$,
while the width of the peaked part grows exponentially with the exponent $\lambda_{0}$. The average momentum and coordinate
in the absence of the bath obey the classical equations of motion. Therefore, the uncertainty of the momentum of a beam decoupled 
from the bath is described by Eqs.~\eqref{ClassSyst1}-\eqref{ClassSyst2} and grows exponentially in time, $\Average{\delta P_z(t)^2}_\prep
\propto e^{2\lambda_{0} t}$. In the presence of the bath, the weight of the peaked part of the distribution gets shifted exponentially
to the smooth part, which results in the additional suppression $\propto e^{-\frac{t}{\tau}}$ of the momentum, with the time variance of the 
uncertainty given by
\begin{align}
		\Average{\delta P_z(t)^2}_\prep
		\propto e^{2\lambda_{0} t-\frac{2t}{\tau}}.
\end{align}

If the distribution of $P_z$ is Gaussian and given by Eq.~\eqref{NormalDitrSingeParticle}], either due to a large number 
of collisions on the time interval considered or due to the initial uncertainty being small and Gaussian, as described in
Sec.~\ref{sec:SingleParticle}, the amount of information stored in the momentum projection is given by
\begin{align}
I_{P}\left(t\right)=\text{const}-\tfrac{1}{2}
\int dP_z \rho\left(P_z,t\right)\ln\left(2\pi\, \delta P_z^2\right)
\nonumber\\
=\text{const}-\lambda t
\label{eq:InfChangeLambdaGeneric}
\end{align}
[the three constants in Eq.~\eqref{eq:def-information} and in the two lines of \eqref{eq:InfChangeLambdaGeneric}
are implied to be different].

In this section, we have demonstrated that the amount of information associated with the projection of momentum 
changes with the rate given by the Lyapunov exponent, i.e. the rate of change of the OTOC of this projection in a generic disordered 
electronic system coupled to a bosonic bath. Since the bosons constituting the bath may mediate interactions,
we expect our result for the relation between the Lyapunov exponent and the rate of change of information to hold
in a generic interacting system as well.


\section{Discussion and outlook\label{sec:Conclusion}}

\label{Sec:conclusion}

In this work, we studied chaotic behaviour in a disordered metal coupled to a bath of neutral 
excitations, such as phonons, magnons or plasmons, with short-range spatial correlations.
 We characterised the chaotic behaviour or the lack thereof
by the out-of-time-order correlators (OTOCs) of the system's momentum, although we expect our results to hold
for rather generic operators, such the density of electrons, currents or magnetisation in a particular volume. 

We have computed the OTOC analytically and demonstrated that
it exhibits exponential decay (non-chaotic behaviour) or exponential growth
(chaotic behaviour) depending on whether or not the inelastic 
scattering rate $1/\tau$ of electrons exceeds the single-particle Lyapunov exponent $\lambda_{0}$ in the system decoupled
from the bath. Our result for the Lyapunov exponent is given by Eq.~\eqref{eq:lya-exp}.
The transition between chaotic and non-chaotic behaviour is analogous to a similar    
transition predicted for a system of electrons with short-range interactions in Ref.~\cite{Syzranov:ChaosTransition}.
This transition may be triggered by changing the strength of the bath, e.g., by tuning 
the temperature of the system.

Furthermore, we have established a relation between the rate of change of information 
and the Lyapunov exponent. If the initial state of the system has a small classical uncertainty,
it leads to an uncertainty of the average values of the observables of the system, which 
grows exponentially during the chaotic evolution of the system, as we have demonstrated in this
paper. A measure of the uncertainty of an observable $A(t)$ is the amount of information given by 
Eq.~\eqref{eq:def-information}. We have demonstrated that the rate of change of the amount of information,
associated with a sufficiently generic observable and
measured in units $(1/\ln 2)$th of a bit,
matches, with the opposite sign, the Lyapunov exponent.

The transition between chaotic and non-chaotic behaviour, discussed here, 
reflects, therefore, in the information stored in the system's observables.
This phenomenon has a simple qualitative interpretation. If dissipation is strong,
an observable relaxes to its equilibrium value and thus becomes more certain; 
the amount of information encoded by this observable grows. In the opposite limit 
of a weak bath, the chaotic evolution of the observable appears random, and its uncertainty
grows.

\begin{figure}[t]
	\label{fig:Dot}
	\centering{}\includegraphics[width=0.75\columnwidth]{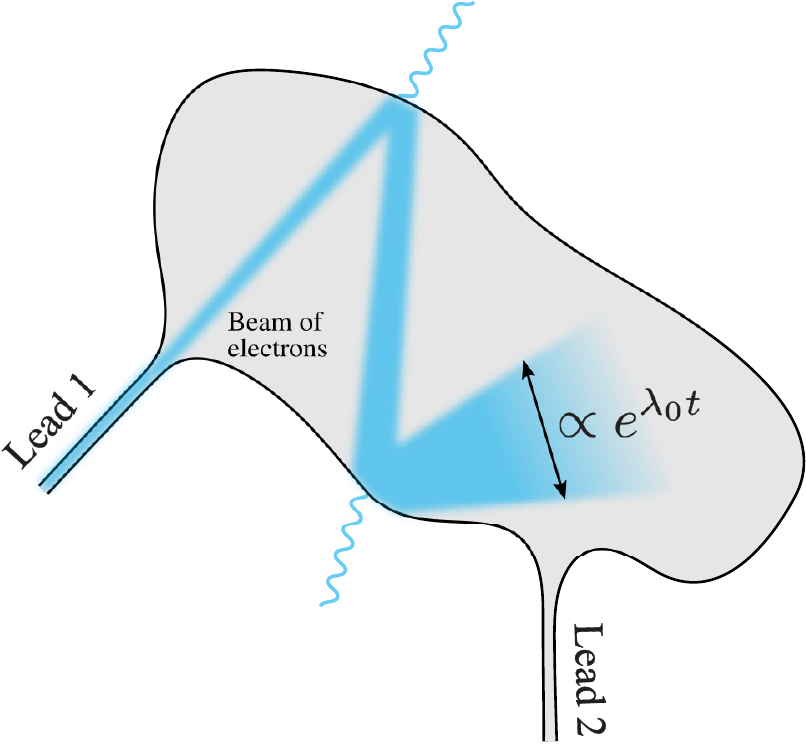}
	\caption{
		(Colour online) A beam of electrons in a chaotic quantum dot. In general,
		the walls of the dot may be partially transparent for the electrons, e.g., in a graphene
		quantum dot~\cite{Velasco:grapheneDot1,Velasco:grapheneDot2}.
		 The escape of the particles through the walls with the rate $1/\tau$ is
		equivalent, for the behaviour of OTOCs, to the effect of a bath with the inelastic scattering
		rate $1/\tau$.
		\label{Fig:dot}
	}
\end{figure}

{\it Measurement.} Our predictions for the Lyapunov exponents for a system of particles in a chaotic billiard
coupled to a dissipative bath and/or in the presence of interactions may be observed straightforwardly, for example,
using trapped ultracold particles whose momenta are excited by a short laser pulse in a small region of space 
and then measuring the distribution of the total momentum, e.g., in a time-of-flight experiment. 

The Lyapunov exponents and their relation to the growth of uncertainty of observables may be measured
in solid-state systems as well, for example, in chaotic quantum dots, as shown in Fig.~\ref{Fig:dot}. The projection $P_z$ of the total 
momentum, whose correlations we considered here, is proportional, for example,
to the magnetic field $B_z$ generated  away from the dot by the current of the electrons in the dot. This
may in principle be used for measuring the correlations of this projections by observing the magnetic noise
when current is injected into the dot through a narrow lead (see Fig.~\ref{Fig:dot}).
The Lyapunov exponent $\lambda=\lambda_0-1/\tau$, computed in this paper, may then 
be observed as a half of the decay rate of the correlator
$\langle [B_z(t+\Delta t)-B_z(t) ]^2\rangle $ where $\langle\ldots\rangle$ is the averaging
over the quantum state of the system and the interval $\Delta t\ll t_E$ may be large,
 but is chosen 
in such a way that the fields $B_z(0)$ and $B_z(\Delta t)$ are close to each other. 
Under this condition, the fields $B_z(t)$ and $B_z(t+\Delta t)$ are equivalent to those occurring from beams of electrons 
corresponding to slightly different initial conditions. We note that the dissipative environment
may in principle be replaced by partially transparent walls of the billiard, for example, in graphene
quantum dots created by electrostatic gating in graphene sheets~\cite{Velasco:grapheneDot1,Velasco:grapheneDot2}, where electrons cannot be confined and
have a finite probability of tunnelling through an arbitrarily high confining potential due to 
Klein tunnelling~\cite{FalkoCheianov,ShytovGuLevitov}. The processes of escape of electrons through 
the walls of the billiard have the same effect on the correlators of observables in the dot as 
a dissipative bath, which reduces the number of particles contributing to the OTOC, as discussed 
in Secs.~\ref{sec:Out-of-time-order-correlator} and \ref{sec:Information}. Furthermore,
the initial width of the beam and the average transparency of the walls of the dot may be tuned electrostatically
by applying gate voltages in graphene quantum dots.

We note that {\it single-particle} chaotic behaviour of electrons has also been predicted to manifest itself
in the shot-noise in quantum dots~\cite{AleinerLarkin:shotNoise}, as well as in all single-particle interference phenomena~\cite{AleinerLarkin,AleinerLarkin:longWL}; these phenomena are expected to display
a crossover between two qualitatively different regimes of behaviour at frequency on the order of
the inverse single-particle Ehrenfest time. While observing this behaviour does not allow one to 
measure the Lyapunov exponents directly, it gives access to the single-particle characteristics of chaos
which affect also the exponents in interacting and/or dissipative systems considered in this paper.
While our arguments here outline generically ways of observing chaos and the dynamic of information
in electronic systems, we leave further detailed analysis of 
specific electronic devices for future work.

{\it Relation to the level statistics.} 
Another aspect of the chaotic phenomena discussed in this paper,
 which ought to be investigated further, is the relation
of Lyapunov exponents and the (non-)existence of chaotic behaviour to the
statistics of the energy levels of the quasiparticles. The equivalence of the classical chaotic behaviour 
to the Wigner-Dyson statistics of the energy levels has been confirmed by vast numerical studies
(see, e.g., Refs.~\cite{Efetov:book,Mirlin:review,Mirlin:reviewSimple} for a review) in 
single-particle systems, such as disordered metals and chaotic billiards.
The many-body levels statistics is also used sometimes to detect phase transitions or quantum
chaotic behaviour, defined by means of OTOCs. However, the many-body level statistics is not directly
accessible in experiment. It still remains to be investigated whether the chaotic behaviour discussed
here and the associated Lyapunov exponents are related to the correlations of the quasiparticle density
of states $\rho(E)=-\frac{1}{\pi V}\int \text{Im} G^R(\br,\br,E)d\br$, where $G^R(\br,\br,E)$ is the retarded 
Green's function in an interacting and/or dissipative system under investigation.


\section{Acknowledgements}

We thank V.~Galitski, E.~Rozenbaum and J.~Schmalian for insightful discussions. 
We are grateful also to V.~Galitski, A.~Gorshkov 
 and J.~Schmalian for prior collaboration
on related topics. MJK's research stay at the University of California Santa Cruz has been supported by the Karlsruhe House of Young Scientists. 
Our work has also been supported by the Hellman Foundation and the 
Faculty Research Grant awarded by the Committee on Research from the University of California, Santa Cruz.


\bibliography{references}

\begin{thebibliography}{50}%
\makeatletter
\providecommand \@ifxundefined [1]{%
 \@ifx{#1\undefined}
}%
\providecommand \@ifnum [1]{%
 \ifnum #1\expandafter \@firstoftwo
 \else \expandafter \@secondoftwo
 \fi
}%
\providecommand \@ifx [1]{%
 \ifx #1\expandafter \@firstoftwo
 \else \expandafter \@secondoftwo
 \fi
}%
\providecommand \natexlab [1]{#1}%
\providecommand \enquote  [1]{``#1''}%
\providecommand \bibnamefont  [1]{#1}%
\providecommand \bibfnamefont [1]{#1}%
\providecommand \citenamefont [1]{#1}%
\providecommand \href@noop [0]{\@secondoftwo}%
\providecommand \href [0]{\begingroup \@sanitize@url \@href}%
\providecommand \@href[1]{\@@startlink{#1}\@@href}%
\providecommand \@@href[1]{\endgroup#1\@@endlink}%
\providecommand \@sanitize@url [0]{\catcode `\\12\catcode `\$12\catcode
  `\&12\catcode `\#12\catcode `\^12\catcode `\_12\catcode `\%12\relax}%
\providecommand \@@startlink[1]{}%
\providecommand \@@endlink[0]{}%
\providecommand \url  [0]{\begingroup\@sanitize@url \@url }%
\providecommand \@url [1]{\endgroup\@href {#1}{\urlprefix }}%
\providecommand \urlprefix  [0]{URL }%
\providecommand \Eprint [0]{\href }%
\providecommand \doibase [0]{http://dx.doi.org/}%
\providecommand \selectlanguage [0]{\@gobble}%
\providecommand \bibinfo  [0]{\@secondoftwo}%
\providecommand \bibfield  [0]{\@secondoftwo}%
\providecommand \translation [1]{[#1]}%
\providecommand \BibitemOpen [0]{}%
\providecommand \bibitemStop [0]{}%
\providecommand \bibitemNoStop [0]{.\EOS\space}%
\providecommand \EOS [0]{\spacefactor3000\relax}%
\providecommand \BibitemShut  [1]{\csname bibitem#1\endcsname}%
\let\auto@bib@innerbib\@empty
\bibitem [{\citenamefont {{Maldacena}}\ \emph {et~al.}(2016)\citenamefont
  {{Maldacena}}, \citenamefont {{Shenker}},\ and\ \citenamefont
  {{Stanford}}}]{Maldacena:bound}%
  \BibitemOpen
  \bibfield  {author} {\bibinfo {author} {\bibfnamefont {J.}~\bibnamefont
  {{Maldacena}}}, \bibinfo {author} {\bibfnamefont {S.~H.}\ \bibnamefont
  {{Shenker}}}, \ and\ \bibinfo {author} {\bibfnamefont {D.}~\bibnamefont
  {{Stanford}}},\ }\href {https://doi.org/10.1007/JHEP08(2016)106} {\bibfield
  {journal} {\bibinfo  {journal} {JHEP}\ }\textbf {\bibinfo {volume} {8}},\
  \bibinfo {pages} {106} (\bibinfo {year} {2016})}\BibitemShut {NoStop}%
\bibitem [{\citenamefont {Larkin}\ and\ \citenamefont
  {Ovchinnikov}(1969)}]{LarkinOvchnnikov}%
  \BibitemOpen
  \bibfield  {author} {\bibinfo {author} {\bibfnamefont {A.}~\bibnamefont
  {Larkin}}\ and\ \bibinfo {author} {\bibfnamefont {Y.~N.}\ \bibnamefont
  {Ovchinnikov}},\ }\href@noop {} {\bibfield  {journal} {\bibinfo  {journal}
  {Sov. Phys. JETP}\ }\textbf {\bibinfo {volume} {28}},\ \bibinfo {pages} {960}
  (\bibinfo {year} {1969})}\BibitemShut {NoStop}%
\bibitem [{\citenamefont {Nakamura}(2004)}]{Nakamura:book}%
  \BibitemOpen
  \bibfield  {author} {\bibinfo {author} {\bibfnamefont {K.}~\bibnamefont
  {Nakamura}},\ }\href@noop {} {\emph {\bibinfo {title} {Quantum Chaos and
  Quantum Dots}}}\ (\bibinfo  {publisher} {Oxford University Press},\ \bibinfo
  {address} {New York},\ \bibinfo {year} {2004})\BibitemShut {NoStop}%
\bibitem [{\citenamefont {Efetov}(1999)}]{Efetov:book}%
  \BibitemOpen
  \bibfield  {author} {\bibinfo {author} {\bibfnamefont {K.~B.}\ \bibnamefont
  {Efetov}},\ }\href@noop {} {\emph {\bibinfo {title} {Supersymetry in Disorder
  and Chaos}}}\ (\bibinfo  {publisher} {Cambridge University Press},\ \bibinfo
  {address} {New York},\ \bibinfo {year} {1999})\BibitemShut {NoStop}%
\bibitem [{\citenamefont {Stanford}(2016)}]{Stanford:firstOTOC}%
  \BibitemOpen
  \bibfield  {author} {\bibinfo {author} {\bibfnamefont {D.}~\bibnamefont
  {Stanford}},\ }\href {\doibase 10.1007/JHEP10(2016)009} {\bibfield  {journal}
  {\bibinfo  {journal} {J. High Energy Phys.}\ }\textbf {\bibinfo {volume}
  {2016}},\ \bibinfo {pages} {9} (\bibinfo {year} {2016})}\BibitemShut
  {NoStop}%
\bibitem [{\citenamefont {{Rozenbaum}}\ \emph {et~al.}(2017)\citenamefont
  {{Rozenbaum}}, \citenamefont {{Ganeshan}},\ and\ \citenamefont
  {{Galitski}}}]{RozenbaumGalitski:rotor}%
  \BibitemOpen
  \bibfield  {author} {\bibinfo {author} {\bibfnamefont {E.~B.}\ \bibnamefont
  {{Rozenbaum}}}, \bibinfo {author} {\bibfnamefont {S.}~\bibnamefont
  {{Ganeshan}}}, \ and\ \bibinfo {author} {\bibfnamefont {V.}~\bibnamefont
  {{Galitski}}},\ }\href {\doibase 10.1103/PhysRevLett.118.086801} {\bibfield
  {journal} {\bibinfo  {journal} {\prl}\ }\textbf {\bibinfo {volume} {118}},\
  \bibinfo {eid} {086801} (\bibinfo {year} {2017})}\BibitemShut {NoStop}%
\bibitem [{\citenamefont {Aleiner}\ \emph {et~al.}(2016)\citenamefont
  {Aleiner}, \citenamefont {Faoro},\ and\ \citenamefont
  {Ioffe}}]{AleinerFaoroIoffe}%
  \BibitemOpen
  \bibfield  {author} {\bibinfo {author} {\bibfnamefont {I.~L.}\ \bibnamefont
  {Aleiner}}, \bibinfo {author} {\bibfnamefont {L.}~\bibnamefont {Faoro}}, \
  and\ \bibinfo {author} {\bibfnamefont {L.~B.}\ \bibnamefont {Ioffe}},\ }\href
  {https://doi.org/10.1016/j.aop.2016.09.006} {\bibfield  {journal} {\bibinfo
  {journal} {Ann. Phys.}\ }\textbf {\bibinfo {volume} {375}},\ \bibinfo {pages}
  {378} (\bibinfo {year} {2016})}\BibitemShut {NoStop}%
\bibitem [{\citenamefont {{Klug}}\ \emph {et~al.}(2017)\citenamefont {{Klug}},
  \citenamefont {{Scheurer}},\ and\ \citenamefont
  {{Schmalian}}}]{KlugSchmalian}%
  \BibitemOpen
  \bibfield  {author} {\bibinfo {author} {\bibfnamefont {M.~J.}\ \bibnamefont
  {{Klug}}}, \bibinfo {author} {\bibfnamefont {M.~S.}\ \bibnamefont
  {{Scheurer}}}, \ and\ \bibinfo {author} {\bibfnamefont {J.}~\bibnamefont
  {{Schmalian}}},\ }\href
  {https://journals.aps.org/prb/abstract/10.1103/PhysRevB.98.045102} {\bibfield
   {journal} {\bibinfo  {journal} {Phys. Rev. B}\ }\textbf {\bibinfo {volume}
  {98}},\ \bibinfo {pages} {045102} (\bibinfo {year} {2017})}\BibitemShut
  {NoStop}%
\bibitem [{\citenamefont {Bagrets}\ \emph {et~al.}(2017)\citenamefont
  {Bagrets}, \citenamefont {Altland},\ and\ \citenamefont
  {Kamenev}}]{BagretsAltlandaKamenev}%
  \BibitemOpen
  \bibfield  {author} {\bibinfo {author} {\bibfnamefont {D.}~\bibnamefont
  {Bagrets}}, \bibinfo {author} {\bibfnamefont {A.}~\bibnamefont {Altland}}, \
  and\ \bibinfo {author} {\bibfnamefont {B.}~\bibnamefont {Kamenev}},\ }\href
  {https://doi.org/10.1016/j.nuclphysb.2017.06.012} {\bibfield  {journal}
  {\bibinfo  {journal} {Nucl. Phys. B}\ }\textbf {\bibinfo {volume} {921}},\
  \bibinfo {pages} {727} (\bibinfo {year} {2017})}\BibitemShut {NoStop}%
\bibitem [{\citenamefont {Patel}\ and\ \citenamefont
  {Sachdev}(2017)}]{PatelSachdev}%
  \BibitemOpen
  \bibfield  {author} {\bibinfo {author} {\bibfnamefont {A.~A.}\ \bibnamefont
  {Patel}}\ and\ \bibinfo {author} {\bibfnamefont {S.}~\bibnamefont
  {Sachdev}},\ }\href {\doibase 10.1073/pnas.1618185114} {\bibfield  {journal}
  {\bibinfo  {journal} {Proc. Natl. Acad. Sci.}\ }\textbf {\bibinfo {volume}
  {114}},\ \bibinfo {pages} {1844} (\bibinfo {year} {2017})}\BibitemShut
  {NoStop}%
\bibitem [{\citenamefont {Patel}\ \emph {et~al.}(2017)\citenamefont {Patel},
  \citenamefont {Chowdhury}, \citenamefont {Sachdev},\ and\ \citenamefont
  {Swingle}}]{PatelChowdhurrySachdevSwingle}%
  \BibitemOpen
  \bibfield  {author} {\bibinfo {author} {\bibfnamefont {A.~A.}\ \bibnamefont
  {Patel}}, \bibinfo {author} {\bibfnamefont {D.}~\bibnamefont {Chowdhury}},
  \bibinfo {author} {\bibfnamefont {S.}~\bibnamefont {Sachdev}}, \ and\
  \bibinfo {author} {\bibfnamefont {B.}~\bibnamefont {Swingle}},\ }\href
  {\doibase 10.1103/PhysRevX.7.031047} {\bibfield  {journal} {\bibinfo
  {journal} {Phys. Rev. X}\ }\textbf {\bibinfo {volume} {7}},\ \bibinfo {pages}
  {031047} (\bibinfo {year} {2017})}\BibitemShut {NoStop}%
\bibitem [{\citenamefont {Werman}\ \emph {et~al.}(2017)\citenamefont {Werman},
  \citenamefont {Kivelson},\ and\ \citenamefont {Berg}}]{WermanKivelsonBerg}%
  \BibitemOpen
  \bibfield  {author} {\bibinfo {author} {\bibfnamefont {Y.}~\bibnamefont
  {Werman}}, \bibinfo {author} {\bibfnamefont {S.~A.}\ \bibnamefont
  {Kivelson}}, \ and\ \bibinfo {author} {\bibfnamefont {E.}~\bibnamefont
  {Berg}},\ }\href {https://arxiv.org/abs/1705.07895} {} (\bibinfo {year}
  {2017}),\ \bibinfo {note} {arXiv:1705.07895}\BibitemShut {NoStop}%
\bibitem [{\citenamefont {{Rozenbaum}}\ \emph {et~al.}(2018)\citenamefont
  {{Rozenbaum}}, \citenamefont {{Ganeshan}},\ and\ \citenamefont
  {{Galitski}}}]{RozenbaumGalitski:statistics}%
  \BibitemOpen
  \bibfield  {author} {\bibinfo {author} {\bibfnamefont {E.~B.}\ \bibnamefont
  {{Rozenbaum}}}, \bibinfo {author} {\bibfnamefont {S.}~\bibnamefont
  {{Ganeshan}}}, \ and\ \bibinfo {author} {\bibfnamefont {V.}~\bibnamefont
  {{Galitski}}},\ }\href@noop {} {\bibfield  {journal} {\bibinfo  {journal}
  {arXiv e-prints}\ ,\ \bibinfo {eid} {arXiv:1801.10591}} (\bibinfo {year}
  {2018})},\ \Eprint {http://arxiv.org/abs/1801.10591} {arXiv:1801.10591
  [cond-mat.dis-nn]} \BibitemShut {NoStop}%
\bibitem [{\citenamefont {Liao}\ and\ \citenamefont
  {Galitski}(2018)}]{LiaoGalitski}%
  \BibitemOpen
  \bibfield  {author} {\bibinfo {author} {\bibfnamefont {Y.}~\bibnamefont
  {Liao}}\ and\ \bibinfo {author} {\bibfnamefont {V.}~\bibnamefont
  {Galitski}},\ }\href {\doibase 10.1103/PhysRevB.98.205124} {\bibfield
  {journal} {\bibinfo  {journal} {Phys. Rev. B}\ }\textbf {\bibinfo {volume}
  {98}},\ \bibinfo {pages} {205124} (\bibinfo {year} {2018})}\BibitemShut
  {NoStop}%
\bibitem [{\citenamefont {{Rozenbaum}}\ \emph {et~al.}(2019)\citenamefont
  {{Rozenbaum}}, \citenamefont {{Bunimovich}},\ and\ \citenamefont
  {{Galitski}}}]{RozenbaumBunimovichGalitski}%
  \BibitemOpen
  \bibfield  {author} {\bibinfo {author} {\bibfnamefont {E.~B.}\ \bibnamefont
  {{Rozenbaum}}}, \bibinfo {author} {\bibfnamefont {L.~A.}\ \bibnamefont
  {{Bunimovich}}}, \ and\ \bibinfo {author} {\bibfnamefont {V.}~\bibnamefont
  {{Galitski}}},\ }\href@noop {} {\bibfield  {journal} {\bibinfo  {journal}
  {arXiv e-prints}\ ,\ \bibinfo {eid} {arXiv:1902.05466}} (\bibinfo {year}
  {2019})},\ \Eprint {http://arxiv.org/abs/1902.05466} {arXiv:1902.05466
  [quant-ph]} \BibitemShut {NoStop}%
\bibitem [{\citenamefont {Syzranov}\ \emph {et~al.}(2019)\citenamefont
  {Syzranov}, \citenamefont {Gorshkov},\ and\ \citenamefont
  {Galitski}}]{Syzranov:ChaosTransition}%
  \BibitemOpen
  \bibfield  {author} {\bibinfo {author} {\bibfnamefont {S.}~\bibnamefont
  {Syzranov}}, \bibinfo {author} {\bibfnamefont {A.}~\bibnamefont {Gorshkov}},
  \ and\ \bibinfo {author} {\bibfnamefont {V.}~\bibnamefont {Galitski}},\
  }\href {\doibase 10.1016/j.aop.2019.03.008} {\bibfield  {journal} {\bibinfo
  {journal} {Annals of Physics}\ }\textbf {\bibinfo {volume} {405}},\ \bibinfo
  {pages} {1 } (\bibinfo {year} {2019})}\BibitemShut {NoStop}%
\bibitem [{\citenamefont {Berman}\ and\ \citenamefont
  {Zaslavsky}(1978)}]{BermanZaslavsky:Ehrenfest}%
  \BibitemOpen
  \bibfield  {author} {\bibinfo {author} {\bibfnamefont {G.}~\bibnamefont
  {Berman}}\ and\ \bibinfo {author} {\bibfnamefont {G.}~\bibnamefont
  {Zaslavsky}},\ }\href {https://doi.org/10.1016/0378-4371(79)90112-2}
  {\bibfield  {journal} {\bibinfo  {journal} {Physica A}\ }\textbf {\bibinfo
  {volume} {91}},\ \bibinfo {pages} {450 } (\bibinfo {year}
  {1978})}\BibitemShut {NoStop}%
\bibitem [{\citenamefont {Slichter}(1996)}]{Slichter:book}%
  \BibitemOpen
  \bibfield  {author} {\bibinfo {author} {\bibfnamefont {C.}~\bibnamefont
  {Slichter}},\ }\href@noop {} {\emph {\bibinfo {title} {Principles of Magnetic
  Resonance}}},\ Springer Series in Solid-State Sciences\ (\bibinfo
  {publisher} {Springer Berlin Heidelberg},\ \bibinfo {year}
  {1996})\BibitemShut {NoStop}%
\bibitem [{\citenamefont {{G{\"a}rttner}}\ \emph {et~al.}(2017)\citenamefont
  {{G{\"a}rttner}}, \citenamefont {{Bohnet}}, \citenamefont {{Safavi-Naini}},
  \citenamefont {{Wall}}, \citenamefont {{Bollinger}},\ and\ \citenamefont
  {{Rey}}}]{GarttnerRey:ionOTOC}%
  \BibitemOpen
  \bibfield  {author} {\bibinfo {author} {\bibfnamefont {M.}~\bibnamefont
  {{G{\"a}rttner}}}, \bibinfo {author} {\bibfnamefont {J.~G.}\ \bibnamefont
  {{Bohnet}}}, \bibinfo {author} {\bibfnamefont {A.}~\bibnamefont
  {{Safavi-Naini}}}, \bibinfo {author} {\bibfnamefont {M.~L.}\ \bibnamefont
  {{Wall}}}, \bibinfo {author} {\bibfnamefont {J.~J.}\ \bibnamefont
  {{Bollinger}}}, \ and\ \bibinfo {author} {\bibfnamefont {A.~M.}\ \bibnamefont
  {{Rey}}},\ }\href {https://doi.org/10.1038/nphys4119} {\bibfield  {journal}
  {\bibinfo  {journal} {Nat. Phys.}\ }\textbf {\bibinfo {volume} {13}},\
  \bibinfo {pages} {781} (\bibinfo {year} {2017})}\BibitemShut {NoStop}%
\bibitem [{\citenamefont {{Bohrdt}}\ \emph {et~al.}(2017)\citenamefont
  {{Bohrdt}}, \citenamefont {{Mendl}}, \citenamefont {{Endres}},\ and\
  \citenamefont {{Knap}}}]{Knap:2copies}%
  \BibitemOpen
  \bibfield  {author} {\bibinfo {author} {\bibfnamefont {A.}~\bibnamefont
  {{Bohrdt}}}, \bibinfo {author} {\bibfnamefont {C.~B.}\ \bibnamefont
  {{Mendl}}}, \bibinfo {author} {\bibfnamefont {M.}~\bibnamefont {{Endres}}}, \
  and\ \bibinfo {author} {\bibfnamefont {M.}~\bibnamefont {{Knap}}},\ }\href
  {https://doi.org/10.1088%2F1367-2630%2Faa719b} {\bibfield  {journal}
  {\bibinfo  {journal} {New J. Phys.}\ }\textbf {\bibinfo {volume} {19}},\
  \bibinfo {pages} {063001} (\bibinfo {year} {2017})}\BibitemShut {NoStop}%
\bibitem [{\citenamefont {{Yao}}\ \emph {et~al.}(2016)\citenamefont {{Yao}},
  \citenamefont {{Grusdt}}, \citenamefont {{Swingle}}, \citenamefont {{Lukin}},
  \citenamefont {{Stamper-Kurn}}, \citenamefont {{Moore}},\ and\ \citenamefont
  {{Demler}}}]{Yao:2copies}%
  \BibitemOpen
  \bibfield  {author} {\bibinfo {author} {\bibfnamefont {N.~Y.}\ \bibnamefont
  {{Yao}}}, \bibinfo {author} {\bibfnamefont {F.}~\bibnamefont {{Grusdt}}},
  \bibinfo {author} {\bibfnamefont {B.}~\bibnamefont {{Swingle}}}, \bibinfo
  {author} {\bibfnamefont {M.~D.}\ \bibnamefont {{Lukin}}}, \bibinfo {author}
  {\bibfnamefont {D.~M.}\ \bibnamefont {{Stamper-Kurn}}}, \bibinfo {author}
  {\bibfnamefont {J.~E.}\ \bibnamefont {{Moore}}}, \ and\ \bibinfo {author}
  {\bibfnamefont {E.~A.}\ \bibnamefont {{Demler}}},\ }\href@noop {} {\bibfield
  {journal} {\bibinfo  {journal} {arXiv e-prints}\ } (\bibinfo {year}
  {2016})},\ \Eprint {http://arxiv.org/abs/1607.01801} {arXiv:1607.01801}
  \BibitemShut {NoStop}%
\bibitem [{\citenamefont {Swingle}\ \emph {et~al.}(2016)\citenamefont
  {Swingle}, \citenamefont {Bentsen}, \citenamefont {Schleier-Smith},\ and\
  \citenamefont {Hayden}}]{Swingle:measurement}%
  \BibitemOpen
  \bibfield  {author} {\bibinfo {author} {\bibfnamefont {B.}~\bibnamefont
  {Swingle}}, \bibinfo {author} {\bibfnamefont {G.}~\bibnamefont {Bentsen}},
  \bibinfo {author} {\bibfnamefont {M.}~\bibnamefont {Schleier-Smith}}, \ and\
  \bibinfo {author} {\bibfnamefont {P.}~\bibnamefont {Hayden}},\ }\href
  {\doibase 10.1103/PhysRevA.94.040302} {\bibfield  {journal} {\bibinfo
  {journal} {Phys. Rev. A}\ }\textbf {\bibinfo {volume} {94}},\ \bibinfo
  {pages} {040302} (\bibinfo {year} {2016})}\BibitemShut {NoStop}%
\bibitem [{\citenamefont {Zhu}\ \emph {et~al.}(2016)\citenamefont {Zhu},
  \citenamefont {Hafezi},\ and\ \citenamefont
  {Grover}}]{ZhuHafeziGrover:measureClock}%
  \BibitemOpen
  \bibfield  {author} {\bibinfo {author} {\bibfnamefont {G.}~\bibnamefont
  {Zhu}}, \bibinfo {author} {\bibfnamefont {M.}~\bibnamefont {Hafezi}}, \ and\
  \bibinfo {author} {\bibfnamefont {T.}~\bibnamefont {Grover}},\ }\href
  {https://doi.org/10.1103/PhysRevA.94.062329} {\bibfield  {journal} {\bibinfo
  {journal} {Phys. Rev. A}\ }\textbf {\bibinfo {volume} {94}},\ \bibinfo
  {pages} {062329} (\bibinfo {year} {2016})}\BibitemShut {NoStop}%
\bibitem [{\citenamefont {{Danshita}}\ \emph {et~al.}(2017)\citenamefont
  {{Danshita}}, \citenamefont {{Hanada}},\ and\ \citenamefont
  {{Tezuka}}}]{Danshita:SYKmeasurement}%
  \BibitemOpen
  \bibfield  {author} {\bibinfo {author} {\bibfnamefont {I.}~\bibnamefont
  {{Danshita}}}, \bibinfo {author} {\bibfnamefont {M.}~\bibnamefont
  {{Hanada}}}, \ and\ \bibinfo {author} {\bibfnamefont {M.}~\bibnamefont
  {{Tezuka}}},\ }\href {https://doi.org/10.1093/ptep/ptx108} {\bibfield
  {journal} {\bibinfo  {journal} {Progr. Theor. Exp. Phys.}\ }\textbf {\bibinfo
  {volume} {2017}},\ \bibinfo {pages} {083I01} (\bibinfo {year}
  {2017})}\BibitemShut {NoStop}%
\bibitem [{\citenamefont {{Li}}\ \emph {et~al.}(2017)\citenamefont {{Li}},
  \citenamefont {{Fan}}, \citenamefont {{Wang}}, \citenamefont {{Ye}},
  \citenamefont {{Zeng}}, \citenamefont {{Zhai}}, \citenamefont {{Peng}},\ and\
  \citenamefont {{Du}}}]{Li:NMRmeas}%
  \BibitemOpen
  \bibfield  {author} {\bibinfo {author} {\bibfnamefont {J.}~\bibnamefont
  {{Li}}}, \bibinfo {author} {\bibfnamefont {R.}~\bibnamefont {{Fan}}},
  \bibinfo {author} {\bibfnamefont {H.}~\bibnamefont {{Wang}}}, \bibinfo
  {author} {\bibfnamefont {B.}~\bibnamefont {{Ye}}}, \bibinfo {author}
  {\bibfnamefont {B.}~\bibnamefont {{Zeng}}}, \bibinfo {author} {\bibfnamefont
  {H.}~\bibnamefont {{Zhai}}}, \bibinfo {author} {\bibfnamefont
  {X.}~\bibnamefont {{Peng}}}, \ and\ \bibinfo {author} {\bibfnamefont
  {J.}~\bibnamefont {{Du}}},\ }\href
  {https://physics.aps.org/featured-article-pdf/10.1103/PhysRevX.7.031011}
  {\bibfield  {journal} {\bibinfo  {journal} {Phys. Rev. X}\ }\textbf {\bibinfo
  {volume} {7}},\ \bibinfo {pages} {031011} (\bibinfo {year}
  {2017})}\BibitemShut {NoStop}%
\bibitem [{\citenamefont {Preskill}(2018)}]{Preskill:book}%
  \BibitemOpen
  \bibfield  {author} {\bibinfo {author} {\bibfnamefont {J.}~\bibnamefont
  {Preskill}},\ }\href
  {http://www.theory.caltech.edu/~preskill/ph219/chap10_6A.pdf} {\enquote
  {\bibinfo {title} {{Quantum Information - Quantum Shannon Theory}},}\ }
  (\bibinfo {year} {2018}),\ \bibinfo {note} {{lecture notes}}\BibitemShut
  {NoStop}%
\bibitem [{Note1()}]{Note1}%
  \BibitemOpen
  \bibinfo {note} {For simplicity, instead of the logarithm to base 2, we
  choose the natural logarithm in the definition of information. The amount of
  information is therefore measured in units of bits times a multiplicative
  constant of ($ 1/\protect \qopname \relax o{ln}2 $)}\BibitemShut {NoStop}%
\bibitem [{\citenamefont {{Basko}}\ \emph {et~al.}(2006)\citenamefont
  {{Basko}}, \citenamefont {{Aleiner}},\ and\ \citenamefont
  {{Altshuler}}}]{BAA}%
  \BibitemOpen
  \bibfield  {author} {\bibinfo {author} {\bibfnamefont {D.~M.}\ \bibnamefont
  {{Basko}}}, \bibinfo {author} {\bibfnamefont {I.~L.}\ \bibnamefont
  {{Aleiner}}}, \ and\ \bibinfo {author} {\bibfnamefont {B.~L.}\ \bibnamefont
  {{Altshuler}}},\ }\href {https://doi.org/10.1016/j.aop.2005.11.014}
  {\bibfield  {journal} {\bibinfo  {journal} {Ann. Phys.}\ }\textbf {\bibinfo
  {volume} {321}},\ \bibinfo {pages} {1126} (\bibinfo {year}
  {2006})}\BibitemShut {NoStop}%
\bibitem [{\citenamefont {Nandkishore}\ and\ \citenamefont
  {Huse}(2015)}]{NandkishoreHuse:review}%
  \BibitemOpen
  \bibfield  {author} {\bibinfo {author} {\bibfnamefont {R.}~\bibnamefont
  {Nandkishore}}\ and\ \bibinfo {author} {\bibfnamefont {D.~A.}\ \bibnamefont
  {Huse}},\ }\href {\doibase 10.1146/annurev-conmatphys-031214-014726}
  {\bibfield  {journal} {\bibinfo  {journal} {Annual Review of Condensed Matter
  Physics}\ }\textbf {\bibinfo {volume} {6}},\ \bibinfo {pages} {15} (\bibinfo
  {year} {2015})},\ \Eprint
  {http://arxiv.org/abs/https://doi.org/10.1146/annurev-conmatphys-031214-014726}
  {https://doi.org/10.1146/annurev-conmatphys-031214-014726} \BibitemShut
  {NoStop}%
\bibitem [{\citenamefont {Altman}\ and\ \citenamefont
  {Vosk}(2015)}]{AltmanVosk:review}%
  \BibitemOpen
  \bibfield  {author} {\bibinfo {author} {\bibfnamefont {E.}~\bibnamefont
  {Altman}}\ and\ \bibinfo {author} {\bibfnamefont {R.}~\bibnamefont {Vosk}},\
  }\href {\doibase 10.1146/annurev-conmatphys-031214-014701} {\bibfield
  {journal} {\bibinfo  {journal} {Annual Review of Condensed Matter Physics}\
  }\textbf {\bibinfo {volume} {6}},\ \bibinfo {pages} {383} (\bibinfo {year}
  {2015})},\ \Eprint
  {http://arxiv.org/abs/https://doi.org/10.1146/annurev-conmatphys-031214-014701}
  {https://doi.org/10.1146/annurev-conmatphys-031214-014701} \BibitemShut
  {NoStop}%
\bibitem [{\citenamefont {{Aleiner}}\ and\ \citenamefont
  {{Larkin}}(1996)}]{AleinerLarkin:longWL}%
  \BibitemOpen
  \bibfield  {author} {\bibinfo {author} {\bibfnamefont {I.~L.}\ \bibnamefont
  {{Aleiner}}}\ and\ \bibinfo {author} {\bibfnamefont {A.~I.}\ \bibnamefont
  {{Larkin}}},\ }\href {https://link.aps.org/doi/10.1103/PhysRevB.54.14423}
  {\bibfield  {journal} {\bibinfo  {journal} {Phys. Rev. B}\ }\textbf {\bibinfo
  {volume} {54}},\ \bibinfo {pages} {14423} (\bibinfo {year}
  {1996})}\BibitemShut {NoStop}%
\bibitem [{\citenamefont {Rammer}\ and\ \citenamefont
  {Smith}(1986)}]{RammerSmith:review}%
  \BibitemOpen
  \bibfield  {author} {\bibinfo {author} {\bibfnamefont {J.}~\bibnamefont
  {Rammer}}\ and\ \bibinfo {author} {\bibfnamefont {H.}~\bibnamefont {Smith}},\
  }\href {https://link.aps.org/doi/10.1103/RevModPhys.58.323} {\bibfield
  {journal} {\bibinfo  {journal} {Rev. Mod. Phys.}\ }\textbf {\bibinfo {volume}
  {58}},\ \bibinfo {pages} {323} (\bibinfo {year} {1986})}\BibitemShut
  {NoStop}%
\bibitem [{\citenamefont {Kamenev}(2011)}]{Kamenev:book}%
  \BibitemOpen
  \bibfield  {author} {\bibinfo {author} {\bibfnamefont {A.}~\bibnamefont
  {Kamenev}},\ }\href@noop {} {\emph {\bibinfo {title} {Field Theory of
  Non-Equilibrium Systems}}}\ (\bibinfo  {publisher} {Cambridge Univ. Press},\
  \bibinfo {address} {Cambridge},\ \bibinfo {year} {2011})\BibitemShut
  {NoStop}%
\bibitem [{\citenamefont {Syzranov}\ \emph {et~al.}(2018)\citenamefont
  {Syzranov}, \citenamefont {Gorshkov},\ and\ \citenamefont
  {Galitski}}]{Syzranov:OTOCdot}%
  \BibitemOpen
  \bibfield  {author} {\bibinfo {author} {\bibfnamefont {S.~V.}\ \bibnamefont
  {Syzranov}}, \bibinfo {author} {\bibfnamefont {A.~V.}\ \bibnamefont
  {Gorshkov}}, \ and\ \bibinfo {author} {\bibfnamefont {V.}~\bibnamefont
  {Galitski}},\ }\href
  {https://journals.aps.org/prb/abstract/10.1103/PhysRevB.97.161114} {\bibfield
   {journal} {\bibinfo  {journal} {Phys. Rev. B}\ }\textbf {\bibinfo {volume}
  {97}},\ \bibinfo {pages} {161114} (\bibinfo {year} {2018})}\BibitemShut
  {NoStop}%
\bibitem [{\citenamefont {Abrikosov}(1988)}]{Abrikosov:metals}%
  \BibitemOpen
  \bibfield  {author} {\bibinfo {author} {\bibfnamefont {A.~A.}\ \bibnamefont
  {Abrikosov}},\ }\href@noop {} {\emph {\bibinfo {title} {Fundamentals of the
  Theory of Metals}}}\ (\bibinfo  {publisher} {Elsevier},\ \bibinfo {address}
  {Oxford},\ \bibinfo {year} {1988})\BibitemShut {NoStop}%
\bibitem [{\citenamefont {{Mahan}}(2000)}]{Mahan:manyParticlePhysics}%
  \BibitemOpen
  \bibfield  {author} {\bibinfo {author} {\bibfnamefont {G.~D.}\ \bibnamefont
  {{Mahan}}},\ }\href@noop {} {\emph {\bibinfo {title} {Many-Particle
  Physics}}}\ (\bibinfo  {publisher} {Springer US},\ \bibinfo {year}
  {2000})\BibitemShut {NoStop}%
\bibitem [{\citenamefont {Zala}\ \emph {et~al.}(2001)\citenamefont {Zala},
  \citenamefont {Narozhny},\ and\ \citenamefont
  {Aleiner}}]{ZalaNarozhnyAleiner}%
  \BibitemOpen
  \bibfield  {author} {\bibinfo {author} {\bibfnamefont {G.}~\bibnamefont
  {Zala}}, \bibinfo {author} {\bibfnamefont {B.~N.}\ \bibnamefont {Narozhny}},
  \ and\ \bibinfo {author} {\bibfnamefont {I.~L.}\ \bibnamefont {Aleiner}},\
  }\href {\doibase 10.1103/PhysRevB.64.214204} {\bibfield  {journal} {\bibinfo
  {journal} {Phys. Rev. B}\ }\textbf {\bibinfo {volume} {64}},\ \bibinfo
  {pages} {214204} (\bibinfo {year} {2001})}\BibitemShut {NoStop}%
\bibitem [{\citenamefont {Gantmakher}(2005)}]{Gantmakher:book}%
  \BibitemOpen
  \bibfield  {author} {\bibinfo {author} {\bibfnamefont {V.~F.}\ \bibnamefont
  {Gantmakher}},\ }\href@noop {} {\emph {\bibinfo {title} {Electrons and
  Disorder in Solids}}}\ (\bibinfo  {publisher} {Oxford University Press},\
  \bibinfo {year} {2005})\BibitemShut {NoStop}%
\bibitem [{\citenamefont {Altshuler}\ and\ \citenamefont
  {Aronov}(1985)}]{AltshulerAronov}%
  \BibitemOpen
  \bibfield  {author} {\bibinfo {author} {\bibfnamefont {B.~L.}\ \bibnamefont
  {Altshuler}}\ and\ \bibinfo {author} {\bibfnamefont {A.~G.}\ \bibnamefont
  {Aronov}},\ }in\ \href@noop {} {\emph {\bibinfo {booktitle}
  {Electron-electron interactions in disordered systems}}}\ (\bibinfo
  {publisher} {North-Holland, Amsterdam},\ \bibinfo {year} {1985})\BibitemShut
  {NoStop}%
\bibitem [{Note2()}]{Note2}%
  \BibitemOpen
  \bibinfo {note} {In the case of a phononic bath, this is realised at
  temperatures exceeding the Debye temperature~\cite
  {Abrikosov:metals}.}\BibitemShut {Stop}%
\bibitem [{\citenamefont {Velasco}\ \emph {et~al.}(2016)\citenamefont
  {Velasco}, \citenamefont {Ju}, \citenamefont {Wong}, \citenamefont {Kahn},
  \citenamefont {Lee}, \citenamefont {Tsai}, \citenamefont {Germany},
  \citenamefont {Wickenburg}, \citenamefont {Lu}, \citenamefont {Taniguchi},
  \citenamefont {Watanabe}, \citenamefont {Zettl}, \citenamefont {Wang},\ and\
  \citenamefont {Crommie}}]{Velasco:grapheneDot1}%
  \BibitemOpen
  \bibfield  {author} {\bibinfo {author} {\bibfnamefont {J.}~\bibnamefont
  {Velasco}}, \bibinfo {author} {\bibfnamefont {L.}~\bibnamefont {Ju}},
  \bibinfo {author} {\bibfnamefont {D.}~\bibnamefont {Wong}}, \bibinfo {author}
  {\bibfnamefont {S.}~\bibnamefont {Kahn}}, \bibinfo {author} {\bibfnamefont
  {J.}~\bibnamefont {Lee}}, \bibinfo {author} {\bibfnamefont {H.-Z.}\
  \bibnamefont {Tsai}}, \bibinfo {author} {\bibfnamefont {C.}~\bibnamefont
  {Germany}}, \bibinfo {author} {\bibfnamefont {S.}~\bibnamefont {Wickenburg}},
  \bibinfo {author} {\bibfnamefont {J.}~\bibnamefont {Lu}}, \bibinfo {author}
  {\bibfnamefont {T.}~\bibnamefont {Taniguchi}}, \bibinfo {author}
  {\bibfnamefont {K.}~\bibnamefont {Watanabe}}, \bibinfo {author}
  {\bibfnamefont {A.}~\bibnamefont {Zettl}}, \bibinfo {author} {\bibfnamefont
  {F.}~\bibnamefont {Wang}}, \ and\ \bibinfo {author} {\bibfnamefont {M.~F.}\
  \bibnamefont {Crommie}},\ }\href {\doibase 10.1021/acs.nanolett.5b04441}
  {\bibfield  {journal} {\bibinfo  {journal} {Nano Letters}\ }\textbf {\bibinfo
  {volume} {16}},\ \bibinfo {pages} {1620} (\bibinfo {year} {2016})},\ \bibinfo
  {note} {pMID: 26852622},\ \Eprint
  {http://arxiv.org/abs/https://doi.org/10.1021/acs.nanolett.5b04441}
  {https://doi.org/10.1021/acs.nanolett.5b04441} \BibitemShut {NoStop}%
\bibitem [{\citenamefont {Lee}\ \emph {et~al.}(2016)\citenamefont {Lee},
  \citenamefont {Wong}, \citenamefont {Velasco~Jr}, \citenamefont
  {Rodriguez-Nieva}, \citenamefont {Kahn}, \citenamefont {Tsai}, \citenamefont
  {Taniguchi}, \citenamefont {Watanabe}, \citenamefont {Zettl}, \citenamefont
  {Wang}, \citenamefont {Levitov},\ and\ \citenamefont
  {Crommie}}]{Velasco:grapheneDot2}%
  \BibitemOpen
  \bibfield  {author} {\bibinfo {author} {\bibfnamefont {J.}~\bibnamefont
  {Lee}}, \bibinfo {author} {\bibfnamefont {D.}~\bibnamefont {Wong}}, \bibinfo
  {author} {\bibfnamefont {J.}~\bibnamefont {Velasco~Jr}}, \bibinfo {author}
  {\bibfnamefont {J.~F.}\ \bibnamefont {Rodriguez-Nieva}}, \bibinfo {author}
  {\bibfnamefont {S.}~\bibnamefont {Kahn}}, \bibinfo {author} {\bibfnamefont
  {H.-Z.}\ \bibnamefont {Tsai}}, \bibinfo {author} {\bibfnamefont
  {T.}~\bibnamefont {Taniguchi}}, \bibinfo {author} {\bibfnamefont
  {K.}~\bibnamefont {Watanabe}}, \bibinfo {author} {\bibfnamefont
  {A.}~\bibnamefont {Zettl}}, \bibinfo {author} {\bibfnamefont
  {F.}~\bibnamefont {Wang}}, \bibinfo {author} {\bibfnamefont {L.~S.}\
  \bibnamefont {Levitov}}, \ and\ \bibinfo {author} {\bibfnamefont {M.~F.}\
  \bibnamefont {Crommie}},\ }\href {\doibase 10.1038/nphys3805
  https://www.nature.com/articles/nphys3805#supplementary-information}
  {\bibfield  {journal} {\bibinfo  {journal} {Nature Physics}\ }\textbf
  {\bibinfo {volume} {12}},\ \bibinfo {pages} {1032} (\bibinfo {year}
  {2016})}\BibitemShut {NoStop}%
\bibitem [{\citenamefont {Cheianov}\ and\ \citenamefont
  {Fal'ko}(2006)}]{FalkoCheianov}%
  \BibitemOpen
  \bibfield  {author} {\bibinfo {author} {\bibfnamefont {V.~V.}\ \bibnamefont
  {Cheianov}}\ and\ \bibinfo {author} {\bibfnamefont {V.~I.}\ \bibnamefont
  {Fal'ko}},\ }\href {\doibase 10.1103/PhysRevB.74.041403} {\bibfield
  {journal} {\bibinfo  {journal} {Phys. Rev. B}\ }\textbf {\bibinfo {volume}
  {74}},\ \bibinfo {pages} {041403} (\bibinfo {year} {2006})}\BibitemShut
  {NoStop}%
\bibitem [{\citenamefont {{Shytov}}\ \emph {et~al.}(2007)\citenamefont
  {{Shytov}}, \citenamefont {{Gu}},\ and\ \citenamefont
  {{Levitov}}}]{ShytovGuLevitov}%
  \BibitemOpen
  \bibfield  {author} {\bibinfo {author} {\bibfnamefont {A.~V.}\ \bibnamefont
  {{Shytov}}}, \bibinfo {author} {\bibfnamefont {N.}~\bibnamefont {{Gu}}}, \
  and\ \bibinfo {author} {\bibfnamefont {L.~S.}\ \bibnamefont {{Levitov}}},\
  }\href@noop {} {\bibfield  {journal} {\bibinfo  {journal} {arXiv e-prints}\
  ,\ \bibinfo {eid} {arXiv:0708.3081}} (\bibinfo {year} {2007})},\ \Eprint
  {http://arxiv.org/abs/0708.3081} {arXiv:0708.3081 [cond-mat.mes-hall]}
  \BibitemShut {NoStop}%
\bibitem [{\citenamefont {{Agam}}\ \emph {et~al.}(2000)\citenamefont {{Agam}},
  \citenamefont {{Aleiner}},\ and\ \citenamefont
  {{Larkin}}}]{AleinerLarkin:shotNoise}%
  \BibitemOpen
  \bibfield  {author} {\bibinfo {author} {\bibfnamefont {O.}~\bibnamefont
  {{Agam}}}, \bibinfo {author} {\bibfnamefont {I.}~\bibnamefont {{Aleiner}}}, \
  and\ \bibinfo {author} {\bibfnamefont {A.}~\bibnamefont {{Larkin}}},\
  }\href@noop {} {\bibfield  {journal} {\bibinfo  {journal} {Phys. Rev. Lett.}\
  }\textbf {\bibinfo {volume} {85}},\ \bibinfo {pages} {3153} (\bibinfo {year}
  {2000})}\BibitemShut {NoStop}%
\bibitem [{\citenamefont {Aleiner}\ and\ \citenamefont
  {Larkin}(1997)}]{AleinerLarkin}%
  \BibitemOpen
  \bibfield  {author} {\bibinfo {author} {\bibfnamefont {I.~L.}\ \bibnamefont
  {Aleiner}}\ and\ \bibinfo {author} {\bibfnamefont {A.~I.}\ \bibnamefont
  {Larkin}},\ }\href {https://link.aps.org/doi/10.1103/PhysRevE.55.R1243}
  {\bibfield  {journal} {\bibinfo  {journal} {Phys. Rev. E}\ }\textbf {\bibinfo
  {volume} {55}},\ \bibinfo {pages} {R1243(R)} (\bibinfo {year}
  {1997})}\BibitemShut {NoStop}%
\bibitem [{\citenamefont {{Mirlin}}(2000)}]{Mirlin:review}%
  \BibitemOpen
  \bibfield  {author} {\bibinfo {author} {\bibfnamefont {A.~D.}\ \bibnamefont
  {{Mirlin}}},\ }\href@noop {} {\bibfield  {journal} {\bibinfo  {journal}
  {Phys. Rep.}\ }\textbf {\bibinfo {volume} {326}},\ \bibinfo {pages} {259}
  (\bibinfo {year} {2000})}\BibitemShut {NoStop}%
\bibitem [{\citenamefont {Mirlin}(2000)}]{Mirlin:reviewSimple}%
  \BibitemOpen
  \bibfield  {author} {\bibinfo {author} {\bibfnamefont {A.~D.}\ \bibnamefont
  {Mirlin}},\ }in\ \href@noop {} {\emph {\bibinfo {booktitle} {Proceedings of
  the International School of Physics ``Enrico Fermi'' on New Directions in
  Quantum Chaos}}},\ \bibinfo {editor} {edited by\ \bibinfo {editor}
  {\bibfnamefont {G.}~\bibnamefont {Casati}}, \bibinfo {editor} {\bibfnamefont
  {I.}~\bibnamefont {Guarneri}}, \ and\ \bibinfo {editor} {\bibfnamefont
  {U.}~\bibnamefont {Smilansky}}}\ (\bibinfo  {publisher} {IOS Press},\
  \bibinfo {address} {Amsterdam},\ \bibinfo {year} {2000})\ pp.\ \bibinfo
  {pages} {223--298}\BibitemShut {NoStop}%
\bibitem [{\citenamefont {Abrikosov}\ \emph {et~al.}(1975)\citenamefont
  {Abrikosov}, \citenamefont {Gorkov},\ and\ \citenamefont
  {Dzyaloshinski}}]{AGD}%
  \BibitemOpen
  \bibfield  {author} {\bibinfo {author} {\bibfnamefont {A.~A.}\ \bibnamefont
  {Abrikosov}}, \bibinfo {author} {\bibfnamefont {L.~P.}\ \bibnamefont
  {Gorkov}}, \ and\ \bibinfo {author} {\bibfnamefont {I.~E.}\ \bibnamefont
  {Dzyaloshinski}},\ }\href@noop {} {\emph {\bibinfo {title} {Methods of
  Quantum Field Theory in Statistical Physics}}}\ (\bibinfo  {publisher}
  {Dover, New York},\ \bibinfo {year} {1975})\BibitemShut {NoStop}%
\bibitem [{\citenamefont {W\"olfle}(2018)}]{Wolfle:Quasiparticles}%
  \BibitemOpen
  \bibfield  {author} {\bibinfo {author} {\bibfnamefont {P.}~\bibnamefont
  {W\"olfle}},\ }\href
  {http://iopscience.iop.org/article/10.1088/1361-6633/aa9bc4/meta} {\bibfield
  {journal} {\bibinfo  {journal} {Rep. Prog. Phys.}\ }\textbf {\bibinfo
  {volume} {81}},\ \bibinfo {pages} {03250} (\bibinfo {year}
  {2018})}\BibitemShut {NoStop}%
\end{thebibliography}%


\onecolumngrid
\newpage

\appendix 

\section{Derivation of the kinetic equation\label{sec:Derivation-kinetic-equation}}

In this Appendix, we provide a detailed derivation of the kinetic
equation Eq.~(\ref{eq:kin-eq}) for the correlator $K\left(\mathbf{R}\mathbf{p},\mathbf{R}'\mathbf{p}',t\right)$,
given by Eqs.~\eqref{eq:def-k}.
The ``ballistic'' part of the equation, i.e. the left-hand side of Eq.~\eqref{eq:kin-eq}, describes the evolution
of the correlator in the presence of the potential $U_{\imp}$, created by the impurities, and in the absence of the phonon bath
and may be derived similarly to the respective part of the conventional kinetic equation~\cite{Mahan:manyParticlePhysics,RammerSmith:review,Abrikosov:metals} for the distribution function $f(\bR,\bp,t)$
of the electrons. In what follows we focus, therefore, on the ``collision integral'', the right-had side of the kinetic equation,
which may be computed under the assumption of a uniform potential.

\begin{figure}[h]
	\begin{centering}
		\includegraphics[width=0.4\paperwidth]{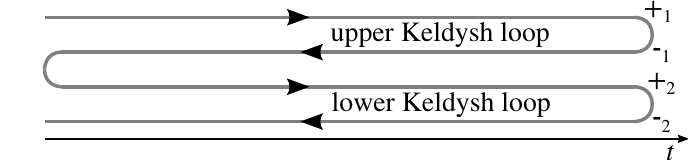}
		\caption{Two-loop Keldysh contour used to compute out-of-time-order correlators.  Branches
			marked with $+_{i}$ and $-_{i}$ mimic propagation forwards and backwards in time on each loop ($i=1$ and $i=2$) of the Keldysh contour.}
		\label{fig:keldyshContour}
		\par\end{centering}
\end{figure}

We study the evolution of the correlation function $K\left(\mathbf{R}\mathbf{p},\mathbf{R}'\mathbf{p}',t\right)$
using Keldysh formalism on the ``augmented'' four-branch Keldysh contour~\cite{AleinerFaoroIoffe} similarly to 
the derivation of the kinetic equation in Ref.~\cite{KlugSchmalian}. To this end, we introduce the averages
\begin{align}
M_{(a,b),c}^{(\alpha,\beta),\gamma}\left(1,2;1',2'\right) & =-\langle\hat{P}_{z}^{c}\left(0\right)\psi^{\dagger b}\left(2'\right)\psi^{a}\left(1'\right)\hat{P}_{z}^{\gamma}\left(0\right)\psi^{\dagger\beta}\left(2\right)\psi^{\alpha}\left(1\right)\rangle_{\mathcal{K}}
\label{eq:Kabc}
\end{align}
of the particle operators ordered on the four-branch Keldysh contour shown in Fig.~\ref{fig:keldyshContour},
where $i=\left(\br_{i},t_{i}\right)$.
In the calculation below, we
use \textit{greek} letters $\alpha,\beta,\gamma\in\left\{ +_{1},-_{1}\right\}$
for operators on the upper branch ($ i=1 $) of the Keldysh contour and  
\textit{latin} letters $a,b,c\in\left\{ +_{2},-_{2}\right\} $ for operators on the lower branch $i=2$
(see Fig.~\ref{fig:keldyshContour}). ``$+_{i}$'' and ``$-_{i}$'' mimic, respectively,
propagation forwards and backwards in time on each loop $i$~\cite{AleinerFaoroIoffe,KlugSchmalian,Stanford:firstOTOC,LiaoGalitski}.
The averages of the form \eqref{eq:Kabc} may be evaluated as path integrals $\langle\dots\rangle_{\mathcal{K}}=\int\mathcal{D}(\psi^{\dagger},\psi,\phi)\dots e^{iS_{\mathcal{K}}}$ 
on the two-loop Keldysh contour, where $\psi^{(\dagger)}$ are the Grassmann
fields representing the electronic degrees of freedom and $\phi$ are 
real-valued fields representing the bosons (phonons) degrees of freedom~\cite{AleinerFaoroIoffe,KlugSchmalian,Kamenev:book}.

Similarly to the case of correlators on the conventional Keldysh contour~\cite{RammerSmith:review,Kamenev:book}, it is possible to introduce various 
correlators of pairs of operators on the two-loop contour in Fig.~\ref{fig:keldyshContour}
with distinct causal properties.
To distinguish between different ways to order a pair of fermionic fields $\psi^\dagger$
and $\psi$ on each loop $i$ in the correlator \eqref{eq:Kabc}, it is convenient also to introduce 
indices  $T \equiv (+_i,+_i )$ (time-ordered), $ \tilde{T} \equiv (-_i,-_i) $ (anti-time-ordered),
$ > \: \equiv (-_i,+_i)$ (greater) and $<\: \equiv (+_i,-_i)$ (lesser) which we use interchangeably
with the labels
``$(\alpha,\beta)$'' and ``$(a,b)$'' in correlators of the form \eqref{eq:Kabc}.
For example, 
the correlator (\ref{eq:def-k}), which reflects the essential properties of the OTOC~\eqref{MomentumOTOC} of the operators of momentum
projections, is given by 
\begin{equation}
K(\mathbf{R}\br\tau,\mathbf{R}'\br'\tau',t)=-\sum_{c,\gamma=\pm}c\gamma M_{<,\gamma}^{<,c}\big(\mathbf{R}\negmedspace+\negmedspace\tfrac{\br}{2}t\negmedspace+\negmedspace\tfrac{\tau}{2},\mathbf{R}\negmedspace-\negmedspace\tfrac{\br}{2}t\negmedspace-\negmedspace\tfrac{\tau}{2};\mathbf{R}'\negmedspace+\negmedspace\tfrac{\br'}{2}t\negmedspace+\negmedspace\tfrac{\tau'}{2},\mathbf{R}'\negmedspace-\negmedspace\tfrac{\br'}{2}t\negmedspace-\negmedspace\tfrac{\tau'}{2}\big)\label{eq:Kapp}
\end{equation}
where $\mathbf{R}^{(\prime)}$ and $\br^{(\prime)}$ are, respectively, the center-of-mass and relative coordinates.

The time evolution of correlators of the form \eqref{eq:Kabc} may be found
 using the perturbation theory on the two-loop Keldysh contour
similarly to the perturbation theory for conventional two-point correlation functions~\cite{AGD,Kamenev:book,Mahan:manyParticlePhysics,RammerSmith:review} and using 
the electron-phonon coupling constant $g$ as a small parameter.
The Dyson equation for the correlation function (\ref{eq:Kabc}) is given, to the leading order in the coupling, by 
\begin{align}
M_{(a,b),c}^{(\alpha,\beta),\gamma} & \left(1,2;1',2'\right)=M_{(a,b),c;0}^{(\alpha,\beta),\gamma}\left(1,2;1',2'\right)\nonumber \\
& +g^{2}\big[\sum_{\delta}\int_{3}\Gamma^{(\alpha,\delta)}(1,3)M_{(a,b),c}^{(\delta,\beta),\gamma}\left(3,2;1',2'\right)+i\sum_{\delta\epsilon}\delta\epsilon\int_{34}G^{(\alpha,\delta)}\left(1,3\right)D^{(\delta,\epsilon)}\left(3,4\right)G^{(\epsilon,\beta)}\left(4,2\right)M_{(a,b),c}^{(\delta,\epsilon),\gamma}\left(3,4;1',2'\right)\nonumber \\
& +\sum_{d}\int_{3}\Gamma^{(a,d)}(1',3')M_{(d,b),c}^{(\alpha,\beta),\gamma}\left(1,2;3',2'\right)+i\sum_{de}de\int_{34}G^{(a,d)}\left(1,3\right)D^{(d,e)}\left(3,4\right)G^{(e,b)}\left(4,2\right)M_{(d,e),c}^{(\alpha,\beta),\gamma}\left(1,2;3,4\right)\nonumber \\
& +i\sum_{\gamma c}d\delta\int_{33'}D^{(\delta,d)}\left(3,3'\right)G^{(a,d)}\left(1',3'\right)G^{(\alpha,\delta)}\left(1,3\right)M_{(d,b),c}^{(\delta,\beta),\gamma}\left(3,2;3',2'\right)\big],
\label{eq:dys1}
\end{align}
where
\begin{align}
	\Gamma^{(\alpha,\beta)}(1,2)=\sum_{\gamma}\int_{3}\gamma\beta G^{(\alpha,\gamma)}(1,3)\Sigma^{(\gamma,\beta)}\left(3,2\right)	
	\label{eq:GammaAppendix}
\end{align}
and 
\begin{equation}
\Sigma^{(\gamma,\beta)}\left(1,2\right)=i G^{(\gamma,\beta)}\left(1,2\right)D^{(\gamma,\beta)}\left(1,2\right)  = \includegraphics[width=0.1\linewidth, valign=c]{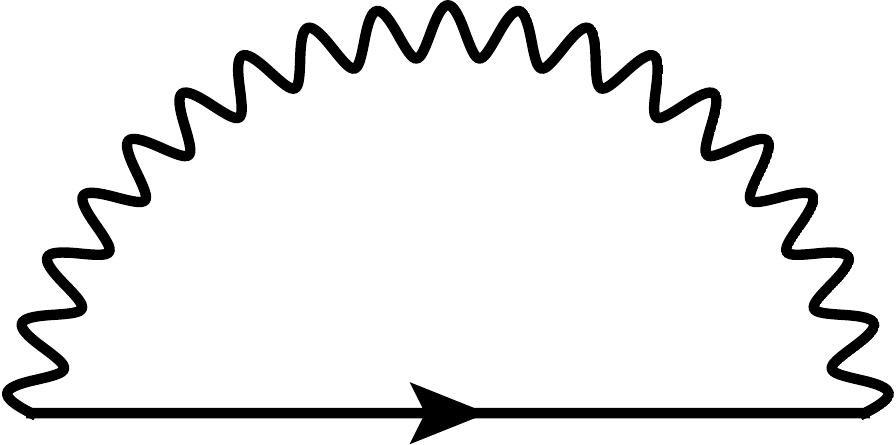}
\label{eq:SigmaAppendix}
\end{equation}
is the electron self-energy with straight and wiggly lines representing, respectively, the fermionic and phononic propagators; the prefactors $\gamma$ and $\beta$ before
the Green's functions in Eq.~\eqref{eq:GammaAppendix} take vales $+1$ and $-1$ for the Green's function
indices ``$ +_i $'' and ``$ -_i $'' on the upper and lower branches of the Keldysh contour on each loop $i$.
In Eqs.~\eqref{eq:dys1}-\eqref{eq:SigmaAppendix} we have also introduced two-point correlation functions
\begin{subequations}
	\label{eq:propagators} 
	\begin{align}
	G^{(\mathsf{a,b})}\left(1,2\right) & =-i\langle\psi^{\mathsf{a}}\left(1\right)\psi^{\dagger\mathsf{b}}\left(2\right)\rangle_{\mathcal{K}_{0}},\\
	D^{(\mathsf{a,b})}\left(1,2\right) & =-i\langle\phi^{\mathsf{a}}\left(1\right)\phi^{\mathsf{b}}\left(2\right)\rangle_{\mathcal{K}_{0}},
	\end{align}
\end{subequations}
where the indices $\mathsf{a,b}\in\{+_{1},-_{1},+_{2},-_{2}\}$ may label any of the four branches 
of the two-loop contour.

Relabelling the indices in Eq.~\eqref{eq:dys1} gives
\begin{align}
M_{(a,b),c}^{(\alpha,\beta),\gamma} & \left(1,2;1',2'\right)=M_{(a,b),c;0}^{(\alpha,\beta),\gamma}\left(1,2;1',2'\right)\nonumber \\
& +g^{2}\big[\sum_{\delta}\int_{3}\Gamma^{(\alpha,\delta)}(3,2)M_{(a,b),c}^{(\delta,\beta),\gamma}\left(1,3;1',2'\right)+i\sum_{\delta\epsilon}\delta\epsilon\int_{34}G^{(\alpha,\delta)}\left(1,3\right)D^{(\delta,\epsilon)}\left(3,4\right)G^{(\epsilon,\beta)}\left(4,2\right)M_{(a,b),c}^{(\delta,\epsilon),\gamma}\left(3,4;1',2'\right)\nonumber \\
& +\sum_{d}\int_{3}\Gamma^{(a,d)}(3',2')M_{(d,b),c}^{(\alpha,\beta),\gamma}\left(1,2;1',3'\right)+i\sum_{de}de\int_{34}G^{(a,d)}\left(1,3\right)D^{(d,e)}\left(3,4\right)G^{(e,b)}\left(4,2\right)M_{(d,e),c}^{(\alpha,\beta),\gamma}\left(1,2;3,4\right)\nonumber \\
& +i\sum_{\gamma c}d\delta\int_{33'}D'^{(\delta,d)}\left(3,3'\right)G^{(d,b)}\left(3',2'\right)G^{(\delta,\beta)}\left(3,2\right)M_{(a,d),c}^{(\alpha,\delta),\gamma}\left(1,3;1',3'\right)\big].
\label{eq:dys2}
\end{align}

\begin{figure}[b]
	\centering
	\includegraphics[width=0.7\linewidth]{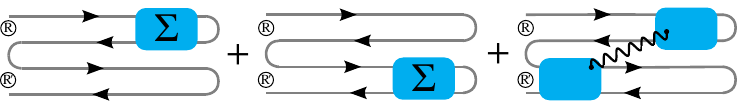}
	\caption{
		(Colour online)
		Leading-order processes which contribute to the self-energy of the four-point correlators
		 of the form \eqref{eq:Kabc}. These correlators account for the correlations
		between electrons with close momenta and with close locations. The first two diagrams describe the conventional self-energies
		of electrons at each location $\bR$ and $\bR^\prime$, respectively. The last diagram
		describes correlations between electrons at these locations.
	}
	\label{fig:contricontour}
\end{figure}

The bosonic propagators may be both interloop or intraloop, i.e. with indices on different loops, e.g. $(\mathsf{a,b}) = (+_{1},-_{2})$,
or with indices on the same loop, e.g. $(\mathsf{a,b}) = (+_{1},-_{1})$.
We note that there are only two distinct types of inter-loop bosonic propagators:
$ D^{(\alpha,+_2)} = D^{(\alpha,-_2)} = D^{(\alpha,a)}\equiv  D_I^{<}  $ and  
$ D^{(a,\alpha)} \equiv  D_I^{>}  $.
The second and third line of Eqs. (\ref{eq:dys1}) and (\ref{eq:dys2}) represent,
respectively, electron-boson scattering processes acting
within one Keldysh loop, whereas the fourth lines denote scattering
processes between the two Keldysh loops.
Below, we introduce the Wigner-transforms of the four-point correlators with the
centre-of-mass coordinates $t^{(\prime)}\negmedspace=\negmedspace\frac{t_{1}^{(\prime)}+t_{2}^{(\prime)}}{2}$ and
 $\mathbf{R}^{(\prime)}\negmedspace=\negmedspace\frac{\br_{1}^{(\prime)}+\br_{2}^{(\prime)}}{2}$ and coordinate differences $\tau^{(\prime)}\negmedspace=\negmedspace t_{1}^{(\prime)}\negmedspace-\negmedspace t_{2}^{(\prime)}$ and $\br^{(\prime)}\negmedspace=\negmedspace\br_{1}^{(\prime)}\negmedspace-\negmedspace\br_{2}^{(\prime)}$.
In this representation, 
intra-loop and the inter-loop processes on the two-loop contour in Fig.~\ref{fig:keldyshContour} correspond, respectively,
to local processes, i.e. processes close to the locations $\bR$ \textit{or} $ \bR' $, and to
the processes which induce correlations between electrons at locations $\bR$
\textit{and} $\bR^\prime$, as shown in Fig.~\ref{fig:contricontour}. 

In what follows, we derive a kinetic equation for the four-point correlator $ K $
similarly to the derivation of the conventional kinetic equation for the distribution function $ f $ (see, e.g., Ref. \cite{Mahan:manyParticlePhysics}). 
To this end, 
we act with the operator $ i\partial _t - \epsilon_{\hat{\bk}}  - U_{\imp}(\br) $, representing single-particle evolution,
on the Dyson's equations  (\ref{eq:dys1}) and (\ref{eq:dys2}), respectively, and use the definition of the single-particle Green's function $[i\partial _{t_1} - \epsilon_{\hat{\bk}_1}  - U_{\imp}(\br_1)-\Sigma^{a,b}(1,2)] G^{(\mathsf{a,b})}\left(1,2\right)  =\mathsf{b}\,\delta\left(1-2\right)\delta_{\mathsf{a,b}} $
to simplify the result. 
By subtracting the obtained expressions and using that $ M_{<}^R = M_{<}^T-M_{<}^< $ and  $ M^{<}_R = M^{<}_T-M^{<}_< $
while introducing the Wigner representation of the correlator $M$ (and, similarly, $K$) at coinciding centre-of-mass times
$t=t^\prime$,
\begin{equation}
M_{(a,b),c}^{(\alpha,\beta),\gamma}(\mathbf{Rp}\omega,\bR'\mathbf{p}'\omega,t)=\int_{\br,\tau}\int_{\br',\tau'}e^{i\omega\tau-i\mathbf{p}\br+i\omega'\tau'-i\mathbf{p}'\br'}M_{(a,b),c}^{(\alpha,\beta),\gamma}(\mathbf{R}\br\tau,\mathbf{R}'\br'\tau',t),
\end{equation}
we obtain the the renormalised retarded components of the correlators in the form 
\begin{subequations}\label{eq:Kr}
	\begin{align}
	M_{<,c}^{R,\gamma}(\mathbf{Rp}\omega,\bR'\mathbf{p}'\omega,t) & =\frac{\mathcal{A}_{<,c}^{\gamma}\left(\mathbf{R}'\mathbf{p}'\omega't\right)}{\omega-\epsilon_{\mathbf{p}}-\Sigma^{R}(\mathbf{p},\omega)},\\
	M_{R,c}^{<,\gamma}(\mathbf{Rp}\omega,\bR'\mathbf{p}'\omega,t) & =\frac{\mathcal{A}_{c}^{<,\gamma}\left(\mathbf{R}\mathbf{p}\omega t\right)}{\omega'-\epsilon_{\mathbf{p}'}-\Sigma^{R}(\mathbf{p}',\omega')},
	\end{align}
\end{subequations}
where $\mathcal{A}_{(a,b),c}^{\gamma}\left(1,2\right)=-\langle\hat{P}_{z}^{c}\left(0\right)\psi^{\dagger b}\left(2\right)\psi^{a}\left(1\right)\hat{P}_{z}^{\gamma}\left(0\right)\rangle_{\mathcal{K}_{0}}$
and $\mathcal{A}_{c}^{(\alpha,\beta),\gamma}(1,2)=-\langle\hat{P}_{z}^{c}\left(0\right)\hat{P}_{z}^{\gamma}\left(0\right)\psi^{\dagger\beta}\left(2\right)\psi^{\alpha}\left(1\right)\rangle_{\mathcal{K}_{0}}$, and have a form similar to that of the retarded two-point propagators.
On the other hand, by adding the two expressions, we obtain the quantum kinetic equation 
\begin{equation}
\label{app:kinEq}
iZ^{-1}(\partial_{t}+i\hat{L}_{\bR,\bp} + i\hat{L}_{\bR',\bp'} )M_{<,c}^{<,\gamma}(\mathbf{Rp}\omega,\bR'\mathbf{p}'\omega,t)= I[M]
\end{equation}
where $i\hat{L}_{\bR,\bp} = \bv_\bp \! \cdot \! \nabla_\bR - \nabla_\bR U_{\imp}(\bR)\! \cdot \! \nabla_\bp  $ is the  the Liouville operator introduced in Eq. (\ref{eq:Liouvillean}) and with the renormalised velocity $\mathbf{v}_{\mathbf{p}}=Z\nabla_{\mathbf{p}}(\epsilon_{\mathbf{p}}+\Sigma^{R}(\mathbf{p},\omega))$;
$Z=(1-\partial_{\omega}\Sigma^{R}(\mathbf{p},\omega))^{-1}$ is the quasiparticle weight; $I[M]$
is the collision integral, which accounts for the relaxation of the correlator due to electron-phonon scattering.
In the weak coupling limit, the renormalisation of the qusiparticle parameters by phonons is weak; hence,
 $  \mathbf{v}_\mathbf{p} \approx v_F \tfrac{\mathbf{p}}{|\mathbf{p}|}$ and $ Z\approx1 $.
When deriving Eqs.~(\ref{eq:Kr}) and (\ref{app:kinEq}) we kept only the leading terms in the gradients
of the centre-of-mass coordinates of the correlators.

Within the collision integral, we distinguish between three types of scattering processes, 
\begin{equation}
I [M] = I_{\text{1}}\left[M\right]+I_{\text{2}}\left[M\right]+I_{\text{3}}\left[M\right], 
\end{equation} 
where the first 
two terms represent local inelastic processes, i.e. occurring at the location 
$ \bR $ \textit{or} $ \bR' $, and the third term leads to correlations between quasiparticles
at the location $\bR$ and those at the location $\bR^\prime$. 
$ I_{\text{1}}\left[M\right]  $ involves correlators with external momenta $ \bp $ and $ \bp' $ and represents the processes of electron scattering from these momentum states to the states with other momenta:
\begin{align}
I_{\text{1}}\left[M\right]= & - g^{2}\big\{\Sigma^{<}(\mathbf{p},\omega)(M_{<,c}^{R,\gamma}-M_{<,c}^{A,\gamma})(\mathbf{Rp}\omega,\bR'\mathbf{p}'\omega,t)+\Sigma^{<}(\mathbf{p}',\omega)(M_{R,c}^{<,\gamma}-M_{A,c}^{<,\gamma})(\mathbf{Rp}\omega,\bR'\mathbf{p}'\omega,t)\nonumber \\
& -\big[ (\Sigma^{>}-\Sigma^{<})(\mathbf{p},\omega)+(\Sigma^{>}-\Sigma^{<})(\mathbf{p}',\omega')\big] M_{<,c}^{<,\gamma}(\mathbf{Rp}\omega,\bR'\mathbf{p}'\omega,t)\big\}. 
\end{align}
Conversely, $ I_2[M] $ contains correlators with internal momenta $ \bq \neq \bp $ (and identically $ \bq' \neq \bp' $). 
It describes, therefore, processes where electrons scatter into the reference state
and is given by  
\begin{align}
I_2 \left[M\right]= & ig^{2}\int_{\mathbf{q}\nu}\big[D^{>}(\mathbf{p}-\mathbf{q},\omega-\nu)G^{<}(\mathbf{R}\mathbf{p}\omega t)(M_{<,c}^{<,\gamma}+M_{<,c}^{R,\gamma}-M_{<,c}^{A,\gamma})(\bR\mathbf{q}\nu,\bR'\mathbf{p}'\omega',t)\nonumber \\
& -G^{>}(\mathbf{R}\bp\omega t)D^{<}(\mathbf{p}-\mathbf{q},\omega-\nu)M_{<,c}^{<,\gamma}\left(\mathbf{R}\mathbf{q}\nu,\mathbf{R}'\mathbf{p}'\omega',t\right)\big]\nonumber \\
& +ig^{2}\int_{\mathbf{q}\nu}\big[D^{>}(\mathbf{p}'-\mathbf{q},\omega'-\nu)G^{<}(\mathbf{R}'\mathbf{p}'\omega't)(M_{<,c}^{<,\gamma}+M_{R,c}^{<,\gamma}-M_{A,c}^{<,\gamma})(\mathbf{R}\mathbf{p}\nu,\mathbf{R}'\mathbf{q}\omega',t)\nonumber \\
& -G^{>}(\mathbf{R}'\mathbf{p}'\omega't)D^{<}(\mathbf{p}'-\mathbf{q},\omega'-\nu)M_{<,c}^{<,\gamma}(\mathbf{R}\mathbf{p}\nu,\mathbf{q}\mathbf{R}'\omega',t)\big].
\end{align}
In contrast to these contributions, accounting for the local processes around $\mathbf{R}$ or $\mathbf{R}'$, 
the contribution $ I_3[M] $ represents processes of quasiparticle scattering between locations $ \bR $ and $ \bR' $ and vice versa. 
It is given by 
\begin{align}
I_{\text{3}} \left[M\right]= & ig^{2}\int_{\mathbf{q}\nu}e^{i\mathbf{q}\left(\mathbf{R-R'}\right)}\big[G^{>}(\mathbf{R}'\mathbf{p}'\negmedspace-\negmedspace\tfrac{\mathbf{q}}{2}\omega'\negmedspace-\negmedspace\tfrac{\nu}{2}t)
D_I^{<}
(\mathbf{q},\nu)M_{<,c}^{<,\gamma}(\mathbf{R}\mathbf{p}\negmedspace-\negmedspace\tfrac{\mathbf{q}}{2}\omega\negmedspace-\negmedspace\tfrac{\nu}{2},\mathbf{R}'\mathbf{p}'\negmedspace+\negmedspace\tfrac{\mathbf{q}}{2}\omega'\negmedspace+\negmedspace\tfrac{\nu}{2},t)\nonumber \\
& -G^{<}(\mathbf{R}'\mathbf{p}'\negmedspace-\negmedspace\tfrac{\mathbf{q}}{2}\omega'\negmedspace-\negmedspace\tfrac{\nu}{2}t)D_I^{<}(\mathbf{q},\nu)(M_{<,c}^{<,\gamma}+M_{<,c}^{R,\gamma}-M_{<,c}^{A,\gamma})(\mathbf{R}\mathbf{p}\negmedspace-\negmedspace\tfrac{\mathbf{q}}{2}\omega\negmedspace-\negmedspace\tfrac{\nu}{2},\mathbf{R}'\mathbf{p}'\negmedspace+\negmedspace\tfrac{\mathbf{q}}{2}\omega'\negmedspace+\negmedspace\tfrac{\nu}{2},t)\big]\nonumber \\
& +ig^{2}\int_{\mathbf{q}\nu}e^{i\mathbf{q}\left(\mathbf{R'-R}\right)}\big[G^{>}(\mathbf{R}\mathbf{p}\negmedspace-\negmedspace\tfrac{\mathbf{q}}{2}\omega\negmedspace-\negmedspace\tfrac{\nu}{2}t)D_I^{>}(\mathbf{q},\nu)M_{<,c}^{<,\gamma}(\mathbf{R}\mathbf{p}\negmedspace+\negmedspace\tfrac{\mathbf{q}}{2}\omega\negmedspace+\negmedspace\tfrac{\nu}{2},\mathbf{R}'\mathbf{p}'\negmedspace-\negmedspace\tfrac{\mathbf{q}}{2}\omega'\negmedspace-\negmedspace\tfrac{\nu}{2},t)\nonumber \\
& -G^{<}(\mathbf{R}\mathbf{p}\negmedspace-\negmedspace\tfrac{\mathbf{q}}{2}\omega\negmedspace-\negmedspace\tfrac{\nu}{2}t)D_I^{>}(\mathbf{q},\nu)(M_{<,c}^{<,\gamma}+M_{R,c}^{<,\gamma}-M_{A,c}^{<,\gamma})(\mathbf{R}\mathbf{p}\negmedspace+\negmedspace\tfrac{\mathbf{q}}{2}\omega\negmedspace+\negmedspace\tfrac{\nu}{2},\mathbf{R}'\mathbf{p}'\negmedspace-\negmedspace\tfrac{\mathbf{q}}{2}\omega'\negmedspace-\negmedspace\tfrac{\nu}{2},t)\big],
\end{align}
We note that the respective matrix element comes with a phase factor which oscillates rapidly as a function of the 
separation $\bR-\bR^\prime$. 
At times $ t\gg t_{\ph} $, where phonon scattering is effectively short-ranged ($|\bR -\bR'| \gg \xi_{\ph}$), 
the contribution $ I_3 $ to the collision integral $I[M]$ is strongly suppressed and, thus, may be neglected. 

In this paper we consider electron dynamics in the regime known as the ``quasiparticle regime''~\cite{Wolfle:Quasiparticles},
where the spectral function $ (G^R-G^A) $, 
as well as $(M_{<,c}^{R,\gamma}-M_{<,c}^{A,\gamma})$ and $(M_{R,c}^{<,\gamma}-M_{A,c}^{<,\gamma})$
[cf. Eqs.~(\ref{eq:Kr})] are sharply peaked at $\omega \approx \epsilon_{\mathbf{p}} $
and contain a negligible incoherent background~\cite{Wolfle:Quasiparticles}.  
It is, therefore, convenient to describe the propagation of quasiparticles by the distribution function
$f\left(\mathbf{R}\mathbf{p}t\right)=-i\int_{\omega}G^{<}\left(\mathbf{R}\mathbf{p}\omega t\right)$.
Similarly, four-point out-of-time-order correlators are characterised conveniently by the function 
\begin{equation}
K\left(\mathbf{R}\mathbf{p},\mathbf{R}'\mathbf{p}',t\right)=-\sum_{c,\gamma=\pm}c\gamma\int_{\omega\omega'}M_{<,\gamma}^{<,c}\left(\mathbf{R}\mathbf{p}\omega t,\mathbf{R}'\mathbf{p}'\omega't\right)
\end{equation}
introduced in Eq. (\ref{eq:Kcl}).

Making these assumptions in Eq. (\ref{app:kinEq}), we obtain the kinetic equation for the correlator
$ K $, which contains also the quasiparticle distribution functions $ f $ whose dynamics are governed by
the conventional kinetic equation~\cite{Abrikosov:metals, RammerSmith:review,Kamenev:book} describing 
charge transport in the system. 
For practical applications, we 
expand all terms in the kinetic equation to the leading order in the deviation of the distribution
function from equilibrium and in the gradients of all functions.
We write $f\left(\mathbf{R}\mathbf{p}t\right)=f_{0}\left(\epsilon_{\mathbf{p}}\right)+g\left(\mathbf{R}\mathbf{p}t\right)$
with the Fermi distribution $f_0(\omega)= \left[\exp(\omega/T)+1\right]^{-1}$
taking into account that $K\left(\mathbf{R}\mathbf{p},\mathbf{R}'\mathbf{p}',t\right)$
vanishes in equilibrium. This gives the kinetic
equation 
\begin{multline}
\label{eq:kinEqApp}
(\partial_{t} +i\hat{L}_{\bR,\bp} + i\hat{L}_{\bR',\bp'} )K\left(\mathbf{R}\mathbf{p},\mathbf{R}'\mathbf{p}',t\right)=\\
-\int_{\mathbf{q}}\left(\Gamma_{\mathbf{p\rightarrow q}}+\Gamma_{\mathbf{p'\rightarrow q}}\right)K\left(\mathbf{R}\mathbf{p},\mathbf{R}'\mathbf{p}',t\right)+\int_{\mathbf{q}}\Gamma_{\mathbf{q\rightarrow p}}K\left(\mathbf{R}\mathbf{q},\mathbf{R}'\mathbf{p}',t\right)+\int_{\mathbf{q}}\Gamma_{\mathbf{q\rightarrow p'}}K\left(\mathbf{R}\mathbf{p},\mathbf{R}'\mathbf{q},t\right),
\end{multline}
where the scattering rates are given by 
	\begin{align}
	\Gamma_{\mathbf{p\rightarrow q}} & =ig^{2}\big[D^{<}\left(\mathbf{p}-\mathbf{q}\right)f_{0}\left(\epsilon_\mathbf{q}\right)+D^{>}\left(\mathbf{p-q}\right)\left(1-f_{0}\left(\epsilon_\mathbf{q}\right)\right)\big], 
	\end{align}
with the greater and lesser boson propagators $D^{<,>}\left(\mathbf{p-q}\right)=D^{<,>}\left(\mathbf{p-q},\epsilon_{\mathbf{p}}-\epsilon_{\mathbf{q}}\right)$.
\end{document}